# How scanning probe microscopy can be supported by Artificial Intelligence and quantum computing?

Agnieszka Pregowska, Agata Roszkiewicz, Magdalena Osial, and Michael Giersig

*Institute of Fundamental Technological Research, Polish Academy of Sciences, Pawinskiego 5B, 02-106 Warsaw, Poland; aprego@ippt.pan.pl (A.P.); arosz@ippt.pan.pl (A.R.); mosial@ippt.pan.pl (M.O.); mgiersig@ippt.pan.pl (M.G.)*

*\* Correspondence: aprego@ippt.pan.pl; Tel.: +48-22-826-1281 (ext. 412)*

*Abstract*—The impact of Artificial Intelligence (AI) is expanding rapidly, revolutionizing both science and society. It is applied to practically all areas of life, science, and technology, including materials science, which continuously needs novel tools for effective materials characterization. One of the widely used techniques is scanning probe microscopy (SPM). SPM has fundamentally changed materials engineering, biology, and chemistry by delivering tools for atomic-precision surface mapping. Besides many advantages, it also has some drawbacks, eg. long scanning time or the possibility of soft-surface materials damage. In this paper, we focus on the potential possibilities for supporting SPM-based measurements, putting emphasis on the application of AI-based algorithms, especially Machine Learning-based algorithms as well as quantum computing (QC). It turned out that AI can be helpful in the experimental processes automation in routine operations, the algorithmic search for good sample regions, and shed light on the structure–property relationships. Thus, it contributes to increasing the efficiency and accuracy of optical nanoscopy scanning probes. Moreover, the combination of AI-based algorithms and QC may have a huge potential to increase the practical application of SPM. The limitations of the AI-QC-based approach were also discussed. Finally, we outline a research path for the improvement of AI-QC-powered SPM.

## I. INTRODUCTION

SCANNING probe microscopy (SPM), including scanning tunneling microscopy (STM), atomic force microscopy (AFM), and scanning near field optical microscopy (SNOM) are universal tools for materials' surface characterization. Such scanning techniques enable the examination of the sample surface with even atomic resolution based on the measurements of the interaction between the probe tip and the sample surface [1]. SPM enables to obtain a high-resolution 3D surface profile in a nondestructive measurement. It can examine samples of various types: metal, dielectric, semiconducting, biological, transparent, etc. without any special preparation [2], usually in air or even liquid conditions, without vacuum requirement. Moreover, it is possible to combine SPM with different techniques to visualize several parameters simultaneously including electrostatic force [3], electronic states [4], ferroelectric domains [5], electric potential [6], magnetic induction [7], adhesion [8], hardness [9], stiffness [10], friction [11], topography [12], chemical structure [13], electronic structure [14], electrochemical reactions [15], local stress [16], impedance [17], resistance [18], electric current flow [19], thermal response [20], optical response [21], polarization [22], refraction index [23], spin angular momentum of electromagnetic fields [24], fluorescence [25], photoluminescence [26], Raman [27] and infrared spectra [28]. Besides its measuring capabilities, SPM can be used to create and modify the structure. It can be done in several ways. First, the simplest techniques rely on mechanically carving patterns in soft material with the use of the AFM probe [29]. The second possibility is to precisely illuminate a photoresist layer through the SNOM probe to create a mask for photolithography [30], [31]. Another technique, called dip-pen nanolithography, uses the probe to absorb the "ink" molecules and release them on the surface during contact mode [32]. Thermal scanning probe lithography is used to directly melt or ablate the substrate [33]. Another technique, field-emission scanning probe lithography, utilizes an electron field emission from the probe material, when a high voltage is applied to the probe [34]. Besides nanolithography, SPM is also a nanomanipulation tool that allows to precise building of desired structures by fine manipulation of small objects (nanoparticles, nanotubes) with the probe tip [35].

SPM faces also some serious challenges (described in more detail in Section IV) connected, among others, with contamination or destruction of the probe or sample during scanning, inaccurate surface mapping due to local defects, tip-sample convolution, slope tilt or overhangs, suboptimal feedback setup, far-field noise, thermal drift, piezoelectric hysteresis, scanner creep, calibration errors. Optical techniques combined with AFM are additionally the source of troublesome optical and chemical artifacts.

Efficient and fast analysis of samples obtained by application of scanning probe microscopy remains a challenge, especially for samples with high attenuation or non-trivial geometries. To understand the spectrum of a sample obtained using SNOM, it is also necessary to specify a wide range of experimental variables such as illumination angle and frequency, the geometry of the tip, and its tapping amplitude. The Artificial Intelligence (AI)-based approach, which already has a wide field of applications, including wastewater treatment [36], [37], circular economy [38], agriculture [39], adsorption studies [40], [41], graphene and graphene-based materials characterization [42], mechanical studies of carbon nanotube-based nanocomposite [43], magnetic hyperthermia [44], optimization of materials properties in cancer diagnosis and treatment [45], early diagnosis of polycystic ovaries syndrome (PCOS) [46], cancer research [47], opened new possibilities in the scanning probe microscopy field [48]. Since AI provides fast extraction of information contained in image data, thus, the application of statistical tools and AI-based algorithms can significantly increase the efficiency and productivity of microscopy, for example, image processing and pattern recognition in scanning transmission electron microscopy (STEM) [49], [50], or enhanced data acquisition and analysis in scanning probe microscopy [51]. AI can be applied to recognition and assigning identities on images, while it enables the detection of specific molecules in complex biological processes enhancing or even replacing hand-engineered features. It is helpful in image segmentation, where the task is to identify whether each pixel belongs to a structure category.

Over the past 70 years, digital computers have made significant progress, from mastering the game of chess to solving complex algebra problems at the school level. The initial enthusiasm for AI peaked with optimistic discussions about household robots that make housework easier and even serve as babysitters. But as media coverage intensified, so did the shadow of skepticism. During these challenges, a promising path has emerged – the convergence of AI and quantum computing (QC) [52]. AI, with its ability to learn complicated tasks, combined with the computing power of quantum computers, offers a unique opportunity for groundbreaking research. This synergy has opened new doors, especially in the processing of huge amounts of data generated during the characterization of nanomaterials using SPM microscopes [53]. The integration of AI and quantum computers allows us to construct learning machines that can distinguish certain properties, such as crystal and electronic structures, from existing electronic data related to nanomaterials. This advancement allows AI to provide valuable insights into potential plasmonic and metamaterial properties, as well as the conditions under which they are generated. As we embark on this exciting journey at the intersection of AI and quantum computing, the possibilities to advance research and overcome previous limitations seem limitless. We are very optimistic about further progress and the transformative impact that this collaboration can bring. The potential possibilities mentioned above are discussed in the following part of this article, using the example of sophisticated SPM microscopes and the digital data generated with their help. The paper is organized as follows: section 2 describes materials and methods, section 3 presents the principle of scanning probe microscopy, section 4 lists the challenges faced by SPM, section 5 presents the theoretical and numerical modeling applied in SPM, section 6 presents next-generation scanning probe microscopy powered by AI, section 7 contains the challenges encountered in SPM supported by AI and QC, while Section 8 contains the discussion, conclusions, and some future remarks.

II. RESEARCH METHODS

We employed a systematic review approach adhering to the PRISMA Statement [54] and its extension, PRISMA-S [55] to systematically evaluate recent publications, reports, protocols, and review papers retrieved from Scopus and Web of Science databases. The data retrieval process involved both electronic and manual searches. The research commenced with a search for relevant research articles to include in the study. The used keywords were: scanning probe microscopy, scanning near-field optical microscopy, scanning tunneling microscopy, atomic force microscope, quantum computing, Artificial Intelligence, Machine Learning, Artificial Neural Networks, and their variations. Based on the research questions:

RQ1: How Artificial Intelligence and data-driven approaches can improve scanning probe microscopy?
RQ2: If convergence of AI and QC has the potential to increase the efficiency of SNOM?

The scope of the study was established, specifying the search period, publication quality, and publication types. Additionally, the chosen sources were scrutinized for alignment with the research topic, and their contribution to nano-spectroscopy was evaluated. Selected texts have been given a certain level of confidence in quality [56]. The search focused on English-language full-text articles, including electronic publications prior to print. Exclusion criteria included doctoral dissertations and materials not related to SPM and AI/ML. After retrieving and analyzing the relevant articles, a total of 252 documents were included in the analysis. The main limitation of the presented study was the fact that in the case of experimental data, only selected data are demonstrated in the literature, and limited access to the data is gained, which can affect the output data. What is more, many different materials are investigated with different techniques, where not only experimental factors should be included as the input but also the type of tools including technical issues. Another issue is the reproducibility of the obtained results, inaccuracy and unreliable selection of input data, incomplete data, and/or complexity in the data.

III. SCANNING PROBE MICROSCOPY – WORKING PRINCIPLE

Scanning probe microscopy is a measurement technique, which allows materials characterization at the atomic level, in particular exploring the local properties of a sample surface with high resolution. It was established in 1982 when Swiss scientists presented the scanning tunneling microscope to investigate a surface with a spatial resolution at the atomic level [57]. This approach is based on the quantum tunneling phenomenon, where a bias voltage is introduced between the surface and a tip, separated by a $0.4 - 0.7$ nm distance in a vacuum. As a result, electrons tunnel through the gap between the tip and the sample, and this tunneling current is a function of the applied voltage and the local electronic density of the states of the sample. The next big step in the history of microscopic measurements was the discovery of the optical near-field scanning microscope [58], [59]. Its operating principle is based on the treatment of the light scattered from a very small gold nanoparticle as a new light source, i.e. utilizing a propagating far field [60]. The excitation and detection of diffraction in the near field are used as microscopic imaging tools, during which the scanning process takes place. According to the Rayleigh criterion, two light lines can be distinguished if the maximum of one diffraction pattern coincides with the minimum of the second one. The resolution of standard light microscopy is limited by the wavelength of light. Subsequent corrections led to achieving a resolution of about 180-200 nm (with the use of immersion oils). SNOM, on the other hand, uses an evanescent near field that carries high-resolution spatial information, but its intensity decreases exponentially with increasing distance from the source. As a consequence, SNOM resolution is limited by the aperture dimensions and probe-sample distance and can be lowered to about 50 nm. Table 1 and Fig. 1 summarize the most important milestones in the evolution of scanning probe microscopy.

TABLE I
THE CRUCIAL MILESTONES IN SCANNING PROBE MICROSCOPY.

| Year | Milestone | The quintessence of discovery | Ref. |
|---|---|---|---|
| 1982 | Scanning tunneling microscope | The bias voltage introduced between the tip and sample surface facilitates electron tunneling across the vacuum gap. Resolution: 0.1 nm (horizontal), 0.01 nm (vertical). | [57] |
| 1984 | Optical near-field scanning microscope | Exploits an evanescent near-field extending beyond the aperture of diameter $< \lambda$. Optical resolution is limited by the aperture diameter. | [58], [59] |
| 1985 | Scanning capacitance microscope | A capacitor is formed between a sharp conducting probe and a semiconductor sample. A bias introduced between the tip and surface of the sample generates capacitance variations. | [61] |
| 1986 | Atomic force microscope | Measures of forces between the mechanical probe and sample. Piezoelectric elements precisely control the movements of the tip. The resolution is even below 1 nm. | [62] |
| 1986 | Scanning thermal microscope | Maps the local temperature and thermal conductivity of an interface with thermal probes (thermocouple, resistive, or bolometer probes). | [63] |
| 1987 | Magnetic force microscope | Utilizes a sharp magnetized tip to probe the magnetic landscape of a sample, enabling the visualization of distinct magnetic domains, including Bloch and Néel walls, closure domains, recorded magnetic bits, and the dynamic behavior of domain walls under the influence of an external magnetic field. | [64] |
| 1987 | Inelastic electron tunneling spectroscopy | Possibility to obtain well-defined vibrational spectra from different parts of the same molecule with the STM probe operating in liquid helium. | [65] |
| 1988 | Electric force microscope | EFM consists of a tip vibrating near a surface and an optical heterodyne detection system to precisely measure its vibrations. Possibility to evaluate tip displacements spanning significant distances and a broad spectrum of frequencies. | [66] |
| 1988 | Inverse photoemission microscope | Detection of light emitted as a result of the inverse photoelectric effect of electrons introduced at a surface through tunneling from a probe. Provides detailed spectroscopic data on the density of unoccupied states and local resonances with an almost atomic resolution across a spectrum of photon energies. Possibility of spatial mapping of optical transitions. | [67] |
| 1989 | Scanning electrochemical microscope | A miniaturized electrode scans over a submerged sample to capture current fluctuations, which are determined by the surface morphology and electrochemical activity of the sample. Possibility to quantify material transfer from a surface with exceptional spatial and temporal precision. | [68] |
| 1989 | Scanning ion-conductance microscope | The tip is an electrode. The non-conducting samples are immersed in aqueous media conducting electrolytes. Measurement of the increase of access resistance in a micropipette when it approaches the surface. | [69] |
| 1989 | Scanning spin-precession microscope | The electron spin precession in a magnetic field generates a fluctuation in the tunneling current, resonating at the Larmor frequency. This radio-frequency signal is confined to regions less than 1 nm in size. Possibility to detect and distinguish individual paramagnetic atoms, spins, and surface defects. | [70] |
| 1990 | Photovoltage scanning tunneling microscope | Photoexcited carriers are drawn towards and accumulate at local potential minima on the surface, leading to enhanced photovoltage. Possibility to investigate both surface and bulk properties, including bandgap variations, doping density changes, and strain field variations. | [71] |
| 1990 | Photon scanning tunneling microscope | The topography of the sample perturbs the surface wave. Photons tunnel between the surface and the probe allowing to measure changes in the evanescent field. No necessity for an electrically conductive surface. | [72] |
| 1991 | Scanning chemical potential microscope | By monitoring the electrical current and thermoelectric voltage generated at the tip-sample junction, this technique offers a direct and sensitive method for assessing atomic-level variations in the surface chemical potential gradient of the heated sample. | [73] |
| 1991 | Kelvin probe force microscope | Measurements of the surface work function contain information about the local composition and electronic state of the surface. The work function is connected to atomic composition, surface defects, catalytic activity, corrosion, trapping of charge in dielectrics, doping and bending of semiconductor bands, phase state, and force distribution on surface reconstruction. | [74] |
| 1994 | Apertureless near-field optical microscope | Detecting the modulation in the electric field of the wave scattered from the sharp tip scanned in immediate contact with the sample surface. Increased resolution of SNOM to the nanometer regime. | [75] |

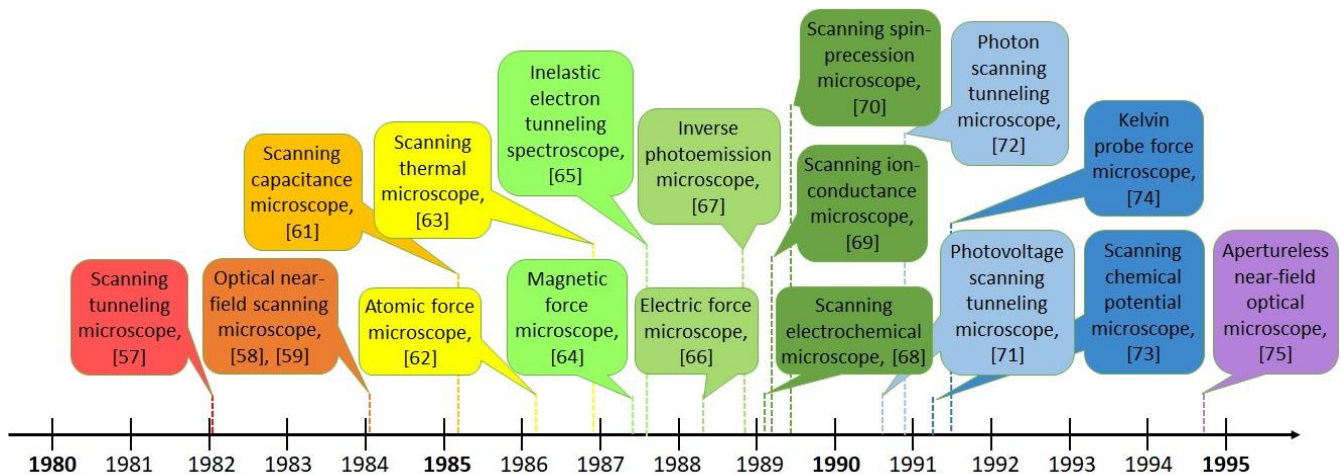

**Fig. 1.** The most important milestones in the development of SPM.

SPM is applied to examine surfaces with atomic-scale resolution. During measurements, a sharp probe is scanned over a sample surface to collect local information concerning the properties of examined materials. The typical separation between the tip and the sample is about 0.1-10 nm. The forces arising between the tip and surface are utilized as a feedback mechanism to regulate the probe-sample distance and bending of the cantilever. The distance and motion of the probe can be controlled through tunneling current assessment, optical deflection control process, fiber interferometry, or piezoresistive techniques [76] The forces that appear between the probe apex and the surface can be classified as attractive, including van der Waals interactions, electrostatic interactions, chemical (covalent bond) interactions and capillary forces (in the presence of a liquid film); or repulsive, including Coulomb interactions between electrons, hard-sphere repulsion (Hertzian contact), and Pauli exclusion repulsion [77]. Each of those interactions is characterized by a different range and strength and their relative influence depends on the tip-sample interval. It is worth noting that the repulsive forces, in general, have a very slight range with inverse power law or exponential decaying dependence of the distance. The resultant complexity of force interaction can make the interpretation of obtained images difficult and ambiguous [78].

The operating modes of the SPM depend on the type of microscope and application. There are four basic modes: contact mode [79] (the tip is constantly in contact with the surface during scanning), tapping mode [80] (the tip vibrates at its resonant frequency or slightly below, gently tapping the sample), non-contact mode [81] (the tip vibrates slightly above its resonant frequency above the surface without touching it) and shear force mode (the cantilever oscillates parallel to the surface) [82]. In addition, the constant mode can be performed under the mechanism of preserving constant force or constant height between the tip and the surface. The real-time feedback system allows controlling the tip-sample distance with high accuracy that could reach even ~ 0.01 Å. In fact, the extensions of the original SPM allowing the gathering of various information about the sample initialized a rapid growth in the field. In combination with the possibility to examine different sample types (metal, semiconductor or dielectric, transparent or opaque) SPM can find its applications in many branches of nanotechnology [83] (nanophotonics, nano-optics, nanolithography), physics [84], chemistry [85], two-dimensional (2D) materials research [86], epitaxial thin-film [87] and electronic circuit components [88], biology [89], climate [90], food [91], pharmacy [92], criminalistics [93], to name a few.

IV. SCANNING PROBE MICROSCOPY – CHALLENGES

Despite its various advantages, SPM also has some disadvantages. A serious limitation is the fact that the information is obtained only from a surface, which automatically excludes a deeper, volume analysis [94]. Another downside is the very small operating tip-sample distance, while the high-resolution images require a long scanning time [95]. Thus, the limited vertical range of a probe restricts the allowed roughness of the examined surface [96]. Another drawback is that the probe is scanned slowly over the surface, resulting in extended scanning time, which may introduce thermal drift in the image. Consequently, this might pose challenges when measuring precise distances between topographical features in the image [97]. As a consequence, a relatively small single scan image size (i.e., maximal about 150×150 μm) is achieved. Another issue is connected with the measurements of the soft materials. Such surfaces examined in a contact or shear force mode might be dragged by the tip or stuck to it, falsifying the outcome [98]. On the other hand, hard materials may damage the probe and thus significantly lower the image quality and lifetime of the probes [99].

Interpretation of the images acquired from SPM requires a detailed analysis of the tip-sample interactions. Optimization of the SPM operation can be very laborious and subject to operator error in terms of sample specifications as well as personal experience

[1]. The multiple types of artifacts can blur the final data and raise doubts about the correctness of the results. The sources of the artifacts in AFM can be various, e.g.: contamination or damage of the sample, damaged, blunt, or contaminated probe tip which may distort the resulting image [100], local defects, tip-sample convolution, slope tilt or overhangs, suboptimal feedback setup, far-field noise, thermal drift, piezoelectric hysteresis, scanner creep, calibration errors. In the case of AFM-based imaging of the biological samples, the generation of lateral forces may contribute to the movement of the sample, or even to its destruction [101]. Moreover, cantilever amplitude during biological sample measurement is sensitive to the sample's structural, mechanical, and chemical properties, leading to potential interference with the accuracy of the measurement [102]. AFM is also inaccurate in the case of measurements of several-layer graphene [103]. The number of possible artifacts increases when AFM is combined with other microscopy techniques like SNOM. The artifacts arising from SNOM are connected with far-field disturbances, shifts between the optical signal and topography, optical contrast (for example interference pattern, artificial stripes, contrast inversion, phase change), or chemical contrast (absorbance, fluorescence, Rayleigh scattering, Raman scattering, polarization, unknown specific shear-force responses), etc. An additional concern arises from the drawbacks of fluorescent probes and slide preparations that involve fluorescent cells. Only in extremely thin samples, these methods do not introduce blurring into the measurements because the fluorescent signals originating beyond the depth of field of the lens are eliminated [104]. All those artifacts may lead to images that significantly differ from an actual surface [105], [106], [107].

V. THE THEORETICAL AND NUMERICAL MODELING IN THE SCANNING PROBE MICROSCOPY

To conduct experiments more effectively, it is crucial to model the process, including the application of Finite-Element Methods (FEM) [108], [109]. For example, the tip geometry in SNOM can be modeled using FEM [110]. It turned out that by modeling the interaction of tip geometry and protective metal coating is possible to refine the experiment before selecting the desired tip configuration. Thus, the tip can be assumed in simple models as a perfect sphere [111]. The more complicated FEM model of SNOM-tip was proposed in [112]. It accounted for the presence of a scatterer in the frequency domain in the form of data-sparse non-local surface-impedance boundary conditions. To reduce computation time the conversion of a sparse finite element (FE) matrix to $\varkappa$-matrix and approximation by adaptive cross approximation algorithm (ACA) to construct the $\varkappa$-matrix have been applied. Numerical methods are also useful when designing apertureless SNOM probes modified with metal or dielectric indentations in order to achieve local near-field enhancement at the apex of the probe [113]. The occurring local geometric resonance, the standing wave resonance of surface plasmon polariton, and the Fano resonance are modeled with the FDTD method. FEM-based methods can be also applied to the modeling of mechanisms for creating nanopatterns on noble metal nanolayers [114]. It was suggested that the main factor to consider when modeling is the melting and transformation of the nanofilm under the tip. FEM can be also applied to the calculation of the real-space electromagnetic field [115], [116], [117], [118]. In [119] artificially created FEM maps were compared to the experimentally inferred spectral shift maps to evaluate the distribution of electric field strength. The optical quality factors as a function of the structural parameters (wall thickness and central hole radius) were calculated from the non-Hermitian perturbation theory [120] These maps can also be obtained by evaluating the total radiating dipole moment of the tip [121], [122]. On the other hand, in the case of AFM measurement FEM-based computation can be used to determine the material loss and surface finish [123] or reproduce the microscopic deformation process [124]. The development of FEM-based methods was proposed in [125], namely the application of the coupled mechanical-electrical finite element approach to the prediction of the reaction of piezoelectric materials to external stimuli. Another approach is the modeling of the material structure and vibrational spectra with the Density Functional Theory (DFT) to evaluate the interaction at the molecular level [126]. DFT can be treated as a complement of experimentally obtained spectra [127], [128]. Thus, the combination of FEM and DFT has the potential to expand the understanding of nanomaterials' structures and their optimization, including surface [129]. For example, in [130] these techniques were combined in the field of scanning Kelvin probe force microscopy analyses of the influence of crystal structure on hydrogen diffusion in α-Fe and γ-Fe.

The modeling of the tip interaction also makes it possible to improve the microscope calibration process, which can be quite long. It can be done in two ways. The first one is computationally expensive and complex, while it requires an approximation of the system of the probe and the sample to an invertible model [131]. The second one is black box calibration models. This approach enables the extraction of the permittivity without detailed electromagnetic interaction modeling. However, the black box method is designed for stationary systems, and the distance between the probe and the surface is slowly modulated, for example in the case of s-SNOM [132], [133]. In [134] calibration method that takes probe tapping into account in extracting the time-invariant sample permittivity was shown. It is based on fitting the Drude model for free electrons to the measured permittivity in the infrared spectral range. In [128] the quantitative s-SNOM model approximates the tip as an elongated conducting spheroid and computes signals from the total radiating dipole moment of the tip. On this basis, parameters such as the Drude weight and electron scattering rate of graphene-supporting propagative plasmons can be obtained.

The numerical methods are also implemented to model physical phenomena that occur during scanning, giving a contribution to the understanding of the experimental data. In [135] the AC–DC module of COMSOL Multiphysics® was used for modeling an electrostatic force-distance curve (EFDC) between an AFM tip and an electrically charged electrode embedded within a thin insulating layer. The study found that the tip and cantilever's contributions to the lateral and vertical potential distribution are significant and must be considered during the interpretation of EFDC measurements. Numerical calculations were also used to explain the underlying mechanisms that contribute to the degradation of asphalt concrete due to moisture at the asphalt-aggregate nanostructured interfaces [136]. The PeakForce Quantitative Nanomechanics AFM was assisted with the Molecular Dynamic simulation (Materials Studio 6.0, Accelrys) based on Lennard-Jone's potential. The calculations revealed the complex structure of the asphalt-aggregate interface and its weak points in the form of water-dissolving asphaltenes and polar aromatics that lead to interface failure. Analytical [137] and numerical [138] models are also used to determine the mechanical properties of soft biological samples that can be characterized only through nanoindentation tests with the use of an AFM tip. The numerical simulations allow also us to determine the reasons for the increased contamination of fluids sterilized by microfiltration when bacteria are treated with sub-lethal concentrations of antibiotics [139]. The AFM nanoindentation measurements revealed that antibiotics reduce the elasticity of bacterial cell membranes and make the walls of bacteria more susceptible to deformation, which leads to increased migration of cells through porous membranes.

VI. NEXT-GENERATION SCANNING PROBE MICROSCOPY POWERED BY ARTIFICIAL INTELLIGENCE AND QUANTUM COMPUTING

Since AI, including Machine Learning (ML) enables efficient analysis of a huge amount of data, it can be considered a powerful computational tool to solve complex problems related to pattern recognition, function estimation, and classification problems [140]. AI techniques also enable inferences about structures that would be difficult to model [141]. Recently, in image processing the deep learning-based approach, i.e. neural networks with a number of hidden layers, has become the dominant methodology in the field of medical image recognition, segmentation [142], and classification [143]. This type of network, based on the input data being a photo or a photo-like image, performs classification and regression tasks. The advantage of deep learning is that it can search the parameter space for the best match of the target and get a solution after the learning phase, in contrast to optimization-based approaches. In the case of scanning probe microscopy, AI-based algorithms can be applied to minimize the need for human action during measurements, and even partially eliminate it [144]. Thus, the automatization of the measurement processes should not be used to eliminate human participation, i.e. autonomous experiment (AE), but since the microscopy experiments are well-defined when we take into account prior physical knowledge, rather than automate routine operations [145]. The schematic idea of the AI/ML-based application in SPM is shown in Fig. 2.

In a deep neural network (DNN) that consists of three layers: an input, an output layer, and at least one hidden layer, artificial neurons are connected with the weight, whose values are selected in the learning process [146]. Hidden layers provide the nonlinear mapping between input and output layers. The effectiveness of these approaches leads to the choice of input data (number and quality), the number of hidden layers, the choice of lost function, the learning rate, and initial weights [147]. DNN can be applied in retrieving subwavelength dimensions based on exclusively far-field data to predict the geometries of the nanostructure, mainly the system of bidirectional network, in which the first one is a geometry-predicting-network (GPN), while the second Spectrum-predicting-network (SPN) [148]. In fact, this is an application of DNNs to solve the inverse problem [149]. A similar approach can be found in the designing of metasurfaces [150], where the inverse problem was formulated as deriving the dielectric function of materials exhibiting subtle or pronounced resonances from experimentally acquired near-field spectra using feed-forward neural networks (FFNNs) proposed. Also, DNN was proposed to retrieve the parameters of a physical object from its scattering pattern with the resolution wavelength/10 [151]. The proposed network architecture consists of four fully connected layers activated with ReLU function and three neurons. The learning algorithm was Adam's stochastic optimization method and the mean absolute error loss function. AI-based solutions can also be applied to the reduction or complete elimination of the time-consuming routine procedures in scanning probe microscopy [152], like, for example, the autonomous SPM operation platform (DeepSPM), which is based on CNNs [153]. It enables assessment of the quality of the acquired images and the condition of the probe.

Thus, the most popular deep learning algorithm used for image processing is the convolutional neural network (CNN) [154]. They contain fewer connections than standard networks with a similar number of hidden values, which makes them easier to train without significantly losing accuracy. This is possible thanks to the operation that allows the flow of information in many planes, i.e. a filter or kernel consisting of a small array of weights [155]. This is beneficial for image analysis. Using the CNNs advantage, which allows to ignoring of irrelevant features in the analysis, such as signal and noise in the ground area, SNOM images can be effectively analyzed [156]. On a millisecond timescale, convolutional neural networks (CNNs) can effectively extract the wavelengths and quality factors of polariton waves from images [157]. In addition, CNNs outperform traditional Fourier transforms in extracting multiple wavelengths for hyperbolic waveguide modes [158]. A 1D supervised CNN can work as a fully automatic technique generating chemical concentration maps from hyperspectral images obtained with Stimulated Raman Scattering [159]. This approach can also be used for the dispersion of quasiparticles [160], evaluation of twist angles, Fermi level lattice parameters, or electron localization degree with Moiré super-lattice imaging [161]. Thus, achieving high CNN accuracy heavily relies on the optimal selection of hire parameters, which is related to the need to provide a large amount of high-quality data [162]. In [163], a CNN was employed to assess the SPM tip quality by analyzing images of known atomic defects on a hydrogen-terminated Si(100) substrate. The proposed network architecture comprises two convolutional layers, one pooling layer, one densely connected layer, and one output layer. It enables automatic identification and isolation subsection of an image obtained using STM. The inputs to the network are fragments containing dangling bonds. On their basis, the network evaluates the quality of the tip. Also, the algorithms based on K nearest neighbor (KNN), Random Forest (RF), Support Vector Machine (SVM), and fully connected neural network (FCNN) with 18 hidden layers with rectified-linear-unit (ReLU) activation function and Adam optimizer was applied for this purpose [164]. The results obtained can be used to develop autonomous atomic-scale production tools. The CNN has been also proposed to be an unattended SPM data acquisition system, namely system DeepSPM [153]. The network is based on 12 convolutional layers and 2 fully connected layers. It was trained with the Adam optimizer [164] with a cross-entropy loss. The system operates as a control loop, scans the sample, and accepts only samples classified as good for analysis. If the sample is classified as bad, it looks for causes such as loss of sample-probe contact, probe failure, error sample region, and invalid probe. In turn, Auto-CO-AFM is an open-source package based on CNNs, which enables the evaluation of the tip functionalization procedures [165].

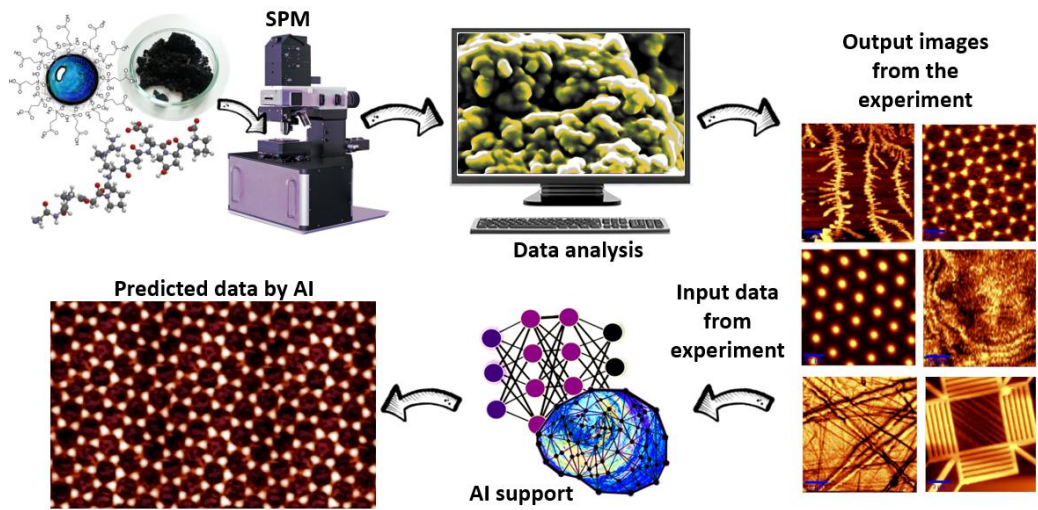

**Fig. 2.** Schematic illustration of AI use in SPM analysis.

Another application of neural networks as a classification tool was shown in [166]. The feed-forward neural network was applied to surface classification based on spin configurations in the case of strongly correlated electronics systems. When the images are near criticality, the spin configuration was obtained with the application of a theoretical mode [167]. The various algorithms can also be combined to provide higher accuracy, like in [168], where the CNN was combined with SVM to differentiate chromosomes into 24 classes. In the first step, to increase the resolution of the input data, the input data was converted using super-resolution models, in particular, the Laplacian pyramid super-resolution network (LapSRN) [169]. Next, the SVM was used to label the input data, and CNN, which consists of six convolution layers, and three pooling layers was applied as a classifier. To increase the accuracy of the network, the Swish activation function was used instead of the ReLU activation function. The intensity profiles perpendicular to the edges, as well as the corresponding edge positions, were the input data to CNN to enhance the microscopic image resolution [170]. Another interesting solution, namely the open-source classification tool for crystalline 2D phases in AFM images (Automated identification of Surface images - AiSurf) was proposed in [171]. It enables the detection of automatically performed such analyses as detecting the distribution of interatomic vectors and deviations from ideal lattices. Also, the use of unsupervised independent component analysis based on non-Gaussianity and statistical independence of data to the Raman spectra of mixtures allows one to extract individual component signals and differentiate organelles in cells by their biochemical compositions without any external labels [172].

On the other hand, in [173] an attempt to reproduce the decision-making process of the person experimenting, namely piezoresponse force microscopy analysis of phase transitions induced by the varied concentrations of Sm dopant in BiFeO3, is shown. This approach combines Gaussian process-based active learning (GP-AL) and Bayesian optimization (GP-BO) procedures. In [174] an indirect adaptive iterative learning control (iAILC) scheme based on iterative dynamic linearization is shown to improve the P-type controller correction response. Learning from setpoint gain is controlled by an adaptive mechanism in real-time. As a result, a linear data model for algorithm design and performance analysis can be obtained for the strongly nonlinear and nonaffine structure of the systems.

AI is also a powerful tool in various types of disease diagnosis. Thus combining AI-based algorithms with SNOM can be useful in the accurate determination of even biological samples, including cancer cells. In [175] multivariate metrics analysis (MA) technique was applied to the precise determination of the system of oral squamous cell carcinoma (OSCC) nodal metastases embedded in the lymphoid tissue with the IR-SNOM (Fourier transform infrared spectroscopy combined with SNOM). The MA-based algorithm can discriminate between two different tissues based on the highest-ranking metric, and as a consequence, provides the possibility of analysis of the chemistry of tissues. In [176], FFNNs were applied to the medical diagnostics of Glioblastoma multiforme (GBM). The summary of different AI-based models and their applications in SPM are presented in Fig. 3.

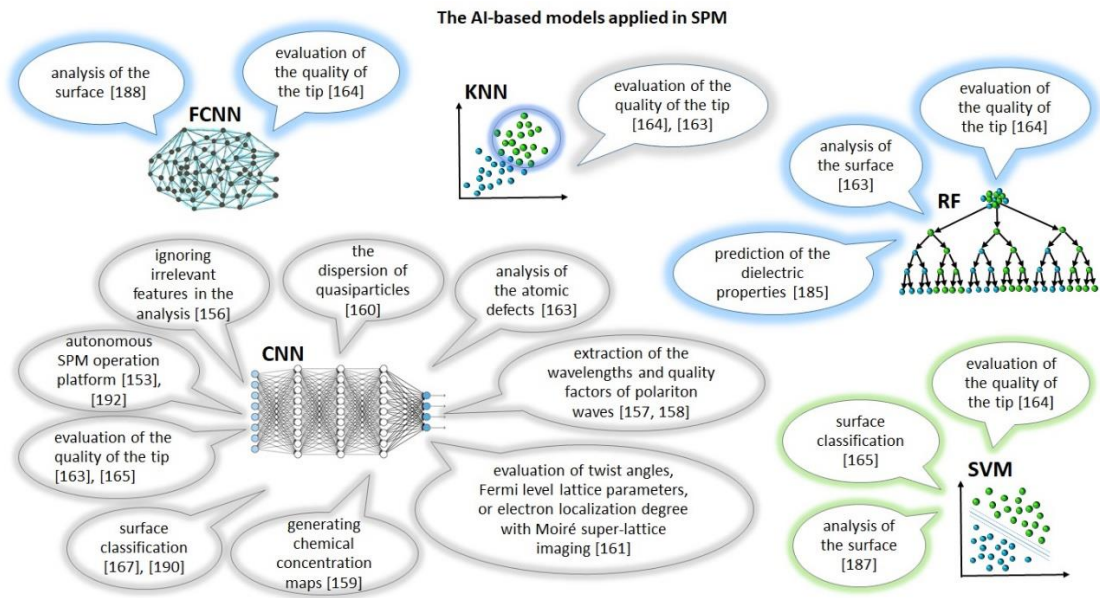

**Fig. 3.** The AI-based models applied in SPM.

An important problem in data analysis is the analysis of noisy data. Here, AI-based approaches are very helpful. This operation can be carried out in many ways, one of them is presented in [177] the application of the deep neural network, which was combined with the least-square approach to extract specific parameters from multidimensional data obtained with the use of spectral-imaging techniques combined with scanning probe (also electron and optical) microscopy. This allows for examination of a wide range of materials for which the excitation is low, and also possibly reduces the need to average signal in time. This approach improves the signal-to-noise ratio of noisy data and enhances pattern recognition, which allows for extracting various material properties from weaker signals.

AI, besides many unquestionable advantages, has some inherent drawbacks. A significant limitation in the development of accurate algorithms is too little data to train them and/or data of poor quality [178]. Thus, data management procedures, in particular, the standards of data annotations with a special emphasis on the Findable, Accessible, Interoperable, Reusable (FAIR) guiding principle are fundamental and are of high importance [179], [180]. In the area of AI, there are two approaches, the first is to work on theoretical (simulated) data like MoleculeNet [181] and the second is the use of experimental measurements like for example the place to store, share, and search in the form of public database SPMImages [182]. An interesting public database was created with the application of Artificial Intelligence, in particular CNN, namely density functional theory (DFT) STM for two-dimensional (2D) materials [183], [184]. It contains data for 716 exfoliated 2D materials, calculated using the Tersoff-Hamann method. Also, by supporting FEM methods, the training and test data sets can be extended with artificially generated data, and as a consequence more efficient algorithms can be developed.

On the other hand, the AI-based algorithm and quantum computation have a huge potential to complement each other and stimulate mutual development [185]. For example, quality control technology stimulates the development of Artificial Intelligence through the control of the parameters optimization. In turn, in solving complex quantum problems an important issue is connected with large-scale quantum devices, namely the effective optimization of parameters of a large number of components, while the complexity of quantum states and dynamics increases potentially with their system size [186]. It is especially important in the context of the SNOM extension to the light frequencies, i.e. the THz spectral region [187]. As a consequence, the analysis of the local electro-optical changes in real space in time and emerging quantitative phases in the THz range is obtained [188]. This solution enables a non-invasive and even almost contactless measurement technique that is dedicated to materials of low-frequency conductivity (in the nanoscale) with high temporal resolution. This leads to the efficient evaluation of material properties like for example, low-energy resonances, or non-dissipative conductivity peaks towards zero frequency [189]. However, Artificial Intelligence uses huge amounts of data for calculations, the processing of which requires significant hardware resources. Thus, quantum Machine Learning (QML) combines AI with quantum computation to develop algorithms for pattern recognition based on the advances of quantum computers like parallel computations and the application of quantum entanglement to perform computations. The first attempt in this field has been made by introducing quantum neural networks (QNNs) [190]. This type of neural network used in computation processes the quantum circuits [191]. In [192] the concept of soft quantum perceptrons (the parameterized quantum circuit with the adjustable parameters of the quantum gate) was introduced. The quantum bits are calculated among others based on the superposition of the state, entanglement, and interference [193]. It turned out that this approach has a nonlinear classification ability compared to the classical perceptrons. The comparison of the performance of AI-based algorithms in SPM is summarized in Table 2 and different AI-based models or tools and flowcharts are presented in Fig. 4.

TABLE II
THE COMPARISON OF THE AI-BASED ALGORITHMS APPLIED IN SCANNING PROBE MICROSCOPY.

| Application field | Accuracy (test sets) | Training/ test sets [%] | Inputs parameters | Outputs parameters | Ref. |
|---|---|---|---|---|---|
| **Algorithm type: RF** | | | | | |
| recognition of the anomalies in the distribution of surface dangling bonds in the hydrogen-terminated silicon surface | 0.89 | 80/20 | 3500 STM images (28x28 pixels) | quality of the tip | [163] |
| prediction of the dielectric properties of polymer nanocomposite interphases | 0.94 | 84/16 | 200 finite-element simulations | interphase permittivity | [194] |
| **Algorithm type: KNN** | | | | | |
| recognition of the anomalies in the distribution of the surface dangling bonds in the hydrogen-terminated silicon surface | 0.84 | 80/20 | 3500 STM images (28x28 pixels) | quality of the tip | [163] |
| **Algorithm type: SVM** | | | | | |
| recognition of the anomalies in the distribution of the surface dangling bonds in the hydrogen-terminated silicon surface | 0.88 | 80/20 | 3500 STM images (28x28 pixels) | quality of the tip | [163] |
| prediction of the dielectric properties of polymer nanocomposite interphases | 0.94 | 84/16 | 200 finite-element simulations | interphase permittivity | [195] |
| prediction of the dielectric properties of polymer nanocomposite interphases | 0.92 | 80/20 | 200 finite-element simulations | interface thickness and permittivity | [196] |
| **Algorithm type: MA** | | | | | |
| cancer diagnostics | 0.99 | 66.67/33.33 | SNOM images | precise determination of oral squamous cell carcinoma | [175] |
| **Algorithm type: FCNN** | | | | | |
| recognition of the anomalies in the distribution of the surface dangling bonds in the hydrogen-terminated silicon surface | 0.78 | 80/20 | 3500 STM images (28x28 pixels) | quality of the tip | [163] |
| segmentation of movable nanowires | 0.90 | about 80/20 | 220 AFM images of silver nanowires (112x112 pixels) | representation of the nanowires in the form of a polygonal line | [197] |
| **Algorithm type: CNN** | | | | | |
| recognition of the anomalies in the distribution of the surface dangling bonds in the hydrogen-terminated silicon surface | 0.97 | 80/20 | 3500 STM images (28x28 pixels) | quality of the tip | [163] |
| recognition of the spatial configurations, identification of the underlying Hamiltonian from a single domain configuration | 0.99 | 80/20 | scattering scanning near-field infrared microscopy (SNIM) images | Hamiltonian | [198] |
| **Algorithm type: CNN** You Only Look Once version 3 (YOLOv3) [186] | | | | | |
| segmentation of movable nanowires | 0.90 | about 80/20 | 220 AFM images of silver nanowires (112x112 pixels) | representation of the nanowires in the form of a polygonal line | [197] |
| generation of STM images of exfoliated 2D materials | 0.90 | 90/10 | 170 black/white images (64 × 64 pixels) | systematic database of STM images obtained using DFT | [199] |
| data acquisition, the algorithmic search of good sample regions, evaluation of the condition of the probe | 0.94 | 76/24 | 7589 constant-current STM topography images (64 × 64 pixels) | condition of the probe, appropriate action location, identification and rectification of probe contact loss/crash incidents | [153], [200] |
| **Algorithm type: CNN+SVM** | | | | | |
| classification of super-resolution enhanced chromosome images | 94.6 | 70/30 | 5474 images of chromosomes (64 × 64 pixels) | classified super-resolution enhanced chromosome | [168], [201] |

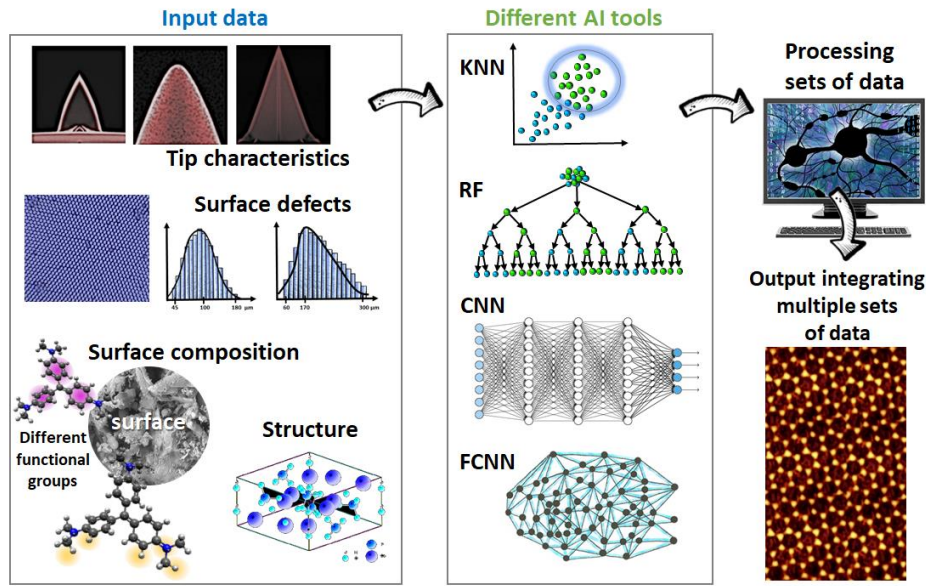

**Fig. 4.** Different AI-based models/tools and flowcharts for SPM.

VII. SCANNING PROBE MICROSCOPY SUPPORTED BY ARTIFICIAL INTELLIGENCE AND QUANTUM COMPUTING - CHALLENGES

Since image acquisition and image analysis are not performed at the same time, a lot of unused data is produced [202]. Additionally, the training data which are used by AI to train is usually prepared manually. This is a time-consuming and laborious task that is not free of human errors [203]. The resulting accuracy and reproducibility can limit confidence in the results' correctness [204]. To reduce human error it is recommended that more than one expert labels training and validation data. Another known problem that occurs during training a machine learning algorithm is overfitting [205], [206]. It happens when the model does not generalize the obtained information and its interpretations are accurate only for the training data. It is a consequence of too small training data size [207], too noisy data [171], underrepresented data [208], too long training on the same dataset [209], or too complex a model that can take noise into account [210]. Strategies to overcome this drawback include increasing the data sets [211], transfer learning, i.e. the knowledge gathered in the previous task is used in the next one [212], cross-validation of the model performance on new data [213], regularization (data augmentation, which consists in modifying the copies of training data to artificially increase the training dataset [202], test-time augmentation, which increases and diverse the test dataset [214], random dropout of selected neurons during training [158, 207], early stopping that ends training when the validation set is no longer improved with updated parameters [215], [216], [217] or L1 and L2 regularizations that add a penalty term to the loss function to create unconstrained problems [218], [219]), model selection [206], [220], prior probability distribution, i.e. assessment of the result done before the experiment [221] or pruning the non-critical sections of the decision tree [222].

Besides, the AI technology complexity limits the possibility to understand and modify algorithms to adapt them to particular requirements [223]. Currently, AI cannot surpass humans in the analysis of incomplete, multi-domain data, which are significantly different from the training data [128]. Hence, due to a trade-off between accuracy and generalization, higher reliability in the case of different tasks is achieved with several specialized algorithms than with one versatile network [224]. The choice of the machine learning algorithm should align with the specific requirements and goals of the project to ensure effective and successful implementation. Selection of an AI-based algorithm for a particular application depends on the quality of data, training dataset size, number of features that need to be recognized, the necessity of human annotation, type of learning algorithm, type of network and their parameters, network topology, model complexity, speed and training time, memory requirements and balance between accuracy and interpretability of the results.

The AI-based method also entails the possibility of making mistakes due to training on historical data, which may be subject to operator error. Moreover, the experimental techniques are not efficient in terms of data set production. In fact, numerical calculations and experimentation increase the amount of data needed to develop accurate algorithms. This leads to another issue in application AI/ML-based methods and is connected with replacing static datasets with processes of active generation of data and employing autonomous systems [225]. However, they are not as risky in the case of SPM as, for example, in the case of autonomous cars.

Another important issue in the application of the AI-based system in the field of SPM in practice is cost and computational time. Thus, the combination of AI and QC has a huge potential to shorten the time needed to train and validate the neural network as well as the optimization processes [226]. Still, this union is challenging [227], for example, for hardware and software incompatibilities. A majority of AI-based algorithms function on specialized graphical processors, while quantum computers are based on qubits that operate at low temperatures. All a joint common programming language is a challenge (the AI-based algorithms are written mostly in Python, while QC is in QASM). Other issues are connected with the low-performance levels of quantum computers and corrections of errors. Moreover, the problem of data security remains. For example, some quantum algorithms although they are faster than classical cryptography methods, are not immune to "harvest now, decrypt later" attacks [228].

VIII. DISCUSSION AND CONCLUSIONS

Scanning probe microscopy is a versatile instrument that dramatically improved the resolution of surface examination up to atomic scale [229], [230]. However, it is not a technique without flaws, the biggest ones include various types of artifacts that are important when interpreting the results obtained. In turn, the recognition of natural images using Artificial Intelligence is very advanced, however, the recognition of microscopic images causes problems because they contain a lot of noise and distortion [231]. A further concern is associated with the fact that microscopic images are in shades of gray. The data is also high resolution, i.e. the data is more complex (larger) than natural image data. Thus, the application of Artificial Intelligence, in particular, deep neural networks in the field of scanning probe microscopy may contribute to a significant increase in the efficiency and accuracy of measurements [202]. It can provide an understanding of the material's structures on a level not yet available to humans [232], [233]. While the interaction between the tip and the surface is approximate with the simplified geometry, the AI has a huge potential to predict the essence of the correlation between the signal and the properties of the sample. In turn, the AI-based algorithms can actively select subsequent scan regions based on the output data and predefined acquisition function and thus contribute to improving the selection of scanned areas [234]. Algorithms should take into account all kinds of factors that can lead to bias in measurements, including differences in tip geometry (length, tip radius, shape, and wear) [235], [236]. Another important issue, when developing the computation tools is proposing a high-accuracy method where there is no need to threshold greyscale images.

The comparison of the AI-based algorithmic performance in the field of SPM taking into account algorithms type, their application field, accuracy, the proportions of training sets to test sets, as well as inputs and outputs parameters, has been done in Table 2. It turned out that AI methods in the areas of the SPM are mostly concentrated after the extraction of structural information. The convolutional neural networks provide high accuracy, i.e. above 90.00 percent in the various types of tasks, including tasks relying on the recognition of the anomalies in the distribution of surface dangling bonds in hydrogen-terminated silicon surface [163], segmentation of movable nanowires [237], classification of super-resolution enhanced chromosome images [168], evaluation of the condition of the probe [153], tip conditioning [238] and even generation of STM images of exfoliate 2D materials [198]. While techniques like K nearest neighbor, Random Forest, and Support Vector Machine enable achieving lower accuracy given the same set of inputs. On the other hand, generative models are also applied to evaluate the correlation between the input data and the output data [239]. On the other hand, one of the limitations of using Artificial Intelligence in AE seems to be that it requires a lot of computing power compared to the processing of typical data and data acquisition times [240], [153], [241]. Thus, automatic image recognition combined with experimental measurements gives benefits such as saving time, minimizing errors, and the ability to analyze data with precision not available to the human eye [242]. Since Artificial Intelligence requires a large amount of data [243], however, experimental measurements are time-consuming, and very often theoretical calculations and simulations based on their results are used to develop data that can enlarge the required databases. Also, platforms and libraries that enable the automatic analysis of images at the atomic level are slowly emerging, such for example Atomvision [244].

One of the future directions of research is the development of generative Artificial Intelligence, which can enable the formulation of answers in the natural language regarding the course of microscopic measurements and samples in the SPM [245]. It also can contribute to the creation of next-generation intelligent laboratories for functional nanomaterials [246]. The first attempt has been made in the fields of AFM automation, in particular, data collection [247], [248]. On the other hand, the development of tools for effective analysis of texts, in particular scientific texts, will also enable the creation of structured databases [249]. Another research line is connected with the development of methods for reliable assessment of measurements carried out by non-specialists. However, the key to improving the overall understanding of the process of scanning probe microscopy and the mechanisms that govern it is to combine the experiment with its numerical simulations and support it with AI-based tools.

Thus, Artificial Intelligence, with its capability of modeling and analyzing the results and comparing them with a constantly growing database of measurements, increases the credibility of the new results obtained with SPM. Nowadays, the possibility to use AI resources has significantly expanded the capabilities of scanning probe devices, improving the reliability of their results by eliminating errors coming from ignorance and experiments. The theoretical output obtained from AI models can be correlated with experimental data and introduced into the device control system to obtain more trustworthy models. Thus, integration of the SPM technique with Artificial Intelligence-based algorithms offers the potential to eliminate most of these measurement technique flaws, especially those associated with artifacts or human factors. This integration significantly enhances the process of sample analysis [250]. It also enables an autonomous system operation to optimize and acquire data without on-site supervision [181]. To summarize, the results obtained indicate that both modeling and the use of Artificial Intelligence-based algorithms, in particular Machine Learning techniques can significantly improve scanning probe microscopy and data interpretation processing [225]. The proposed approach can be applied in molecular diagnostics and screening applications. The simulations can offer an understanding of the fundamental physical and chemical principles governing the behavior of materials. In turn, AI-based methods can enable probabilistic learning of atomic interactions. In addition, natural language processing (NLP) can also help organize and classify knowledge in the context of SPM by searching for key information in the literature. Moreover, the combination of Artificial Intelligence and quantum computation may be a benefit, however, both require significant improvement [251], [252]. Thus, the tensor network systems that transmit and process a huge amount of information seem to be implemented in the future in almost every area of life.

## References

[1] K. Bian, C. Gerber, A. J. Heinrich, D. J. Müller, S. Scheuring, Y. Jiang, "Scanning probe microscopy," *Nat. Rev. Methods Primers* vol. 1, pp. 36, 2021, doi: 10.1038/s43586-021-00033-2.

[2] Z. Lyu, L. Yao, W. Chen, F. C. Kalutantirige, C. Chen, "Electron Microscopy Studies of Soft Nanomaterials," *Chem. Rev.* vol. 123, no. 7, pp. 4051-4145, 2023, doi: 10.1021/acs.chemrev.2c00461.

[3] S. Sahare, P. Ghoderao, Y. Chan, S. L. Lee, "Surface supramolecular assemblies tailored by chemical/physical and synergistic stimuli: a scanning tunneling microscopy study," *Nanoscale* vol. 15, pp. 1981-2002, 2023, doi: 10.1039/D2NR05264D.

[4] M. V. Ale Crivillero, J. C. Souza, V. Hasse, M. Schmidt, N. Shitsevalova, S. Gabáni, K. Siemensmeyer, K. Flachbart, S. Wirth, "Detection of Surface States in Quantum Materials $ZrTe_2$ and $TmB_4$ by Scanning Tunneling Microscopy," *Condens. Matter* vol. 8, pp. 9, 2023.

[5] W. S. Huxter, M. F. Sarott, M. Trassin, C. L. Degen, "Imaging ferroelectric domains with a single-spin scanning quantum sensor," *Nat. Phys.* vol. 19, pp. 644-648, 2023, doi: 10.1038/s41567-022-01921-4.

[6] K. Iwaya, M. Yokota, H. Hanada, H. Mogi, S. Yoshida, O. Takeuchi, U. Miyatake, H. Shigekawa, "Externally-triggerable optical pump-probe scanning tunneling microscopy with a time resolution of tens-picosecond," *Sci. Rep.* vol. 13, pp. 818, 2023, doi: 10.1038/s41598-023-27383-z.

[7] T. Johnsen, C. Schattauer, S. Samaddar, A. Weston, M. J. Hamer, K. Watanabe, T. Taniguchi, R. Gorbachev, F. Libisch, K. Morgenstern, "Mapping quantum Hall edge states in graphene by scanning tunneling microscopy," *Phys. Rev. B* vol. 107, pp. 115426, 2003, doi: 10.1103/PhysRevB.107.115426.

[8] S. S. Hassani, M. Daraee, Z. Sobat, "Application of atomic force microscopy in adhesion force measurements," *J. Adhes. Sci. Tech.* vol. 35, pp. 3221-241, 2021, doi: 10.1080/01694243.2020.1798647.

[9] J. Chen, M. Y. Hu, L. Qing, P. Liu, L. Li, R. Li, C. X. Yue, J. H. Lin, "Study on Boundary Layer and Surface Hardness of Carbon Black in Natural Rubber Using Atomic Force Microscopy," *Polymers* vol. 14, no. 21, pp. 4642, 2022. doi: 10.3390/polym14214642.

[10] C. Petit, A. A. Karkhaneh Yousefi, M. Guilbot, V. Barnier, S. Avril, "Atomic Force Microscopy Stiffness Mapping in Human Aortic Smooth Muscle Cells," *J. Biomech. Engineering* vol. 144, no. 8, pp. 081001, 2022, doi: 10.1115/1.4053657.

[11] A. J. Weymouth, O. Gretz, E. Riegel, F. J. Giessibl, "Measuring sliding friction at the atomic scale," *Japan J. Appl. Physics* vol. 61, pp. SL0801, 2022, doi: 10.35848/1347-4065/ac5e4a.

[12] R. Cojocaru, O. Mannix, M. Capron, C. G. Miller, P. H. Jouneau, B. Gallet, D. Falconet, A. Pacureanu, S. Stukins, "A biological nanofoam: The wall of coniferous bisaccate pollen," *Sci. Adv.* vol. 8, no. 6, pp. eabd0892, 2022, doi: 10.1126/sciadv.abd0892.

[13] M. Commodo, K. Kaiser, G. De Falco, P. Minutolo, F. Schulz, A. D'Anna, L. Gross, "On the early stages of soot formation: Molecular structure elucidation by high-resolution atomic force microscopy," *Combust Flame* vol. 205, pp. 154-164, 2019, doi: 10.1016/j.combustflame.2019.03.042.

[14] Y. Yuan, X. Wang, H. Li, J. Li, Y. Ji, Z. Hao, Y. Wu, K. He, Y. Wang, Y. Xu, W. Duan, W. Li W, Q. K. Xu, "Electronic states and magnetic response of $MnBi_2Te_4$ by scanning tunneling microscopy and spectroscopy," *Nano Letters* vol. 20, no. 5, pp. 3271-3277, 2020, doi: 10.1021/acs.nanolett.0c00031.

[15] A. Asserghine, A. M. Ashrafi, A. Mukherjee, F. Petrlak, Z. Heger, P. Svec, L. Richtera, L. Nagy, R. M. Souto, G. Nagy, V. Adam, "In situ investigation of the cytotoxic and interfacial characteristics of titanium when galvanically coupled with magnesium using scanning electrochemical microscopy," *ACS Appl. Mater. Interfaces* vol. 13, no. 36, pp. 43587-43596, 2021, doi: 10.1021/acsami.1c10584.

[16] L. Haonan, X. Liang, K. Nakajima, "Nanoscale strain–stress mapping for a thermoplastic elastomer revealed using a combination of in situ atomic force microscopy nanomechanics and Delaunay triangulation," *J. Polymer Sci.* vol. 6022, pp. 3134-3140, 2022, doi: 10.1002/pol.20220345.

[17] V. Shkirskiy, M. Kang, I. J. McPherson, C. L. Bentley, O. J. Wahab, E. Daviddi, A. W. Colburn, P. R. Unwin, "Electrochemical impedance measurements in scanning ion conductance microscopy," *Anal. Chem.* vol. 92, no. 18, pp. 12509-12517, 2020, doi: 10.1021/acs.analchem.0c02358.

[18] M. Waldrip, O. D. Jurchescu, D. J. Gundlach, E. G. Bittle, "Contact resistance in organic field-effect transistors: conquering the barrier," *Adv. Funct. Mater.* vol. 30, no. 20, pp. 1904576, 2020, doi: 10.1002/adfm.201904576.

[19] R. Giridharagopal, J. T. Precht, S. Jariwala, L. Collins, S. Jesse, S. V. Kalinin, D. S. Ginger, "Time-resolved electrical scanning probe microscopy of layered perovskites reveals spatial variations in photoinduced ionic and electronic carrier motion,' *ACS Nano* vol. 13, no. 3, pp. 2812-2821, 2019, doi: 10.1021/acsnano.8b08390.

[20] S. Vaziri, E. Yalon, M. Muñoz Rojo, S. V. Suryavanshi, H. Zhang, C. J. McClellan, C. S. Bailey, K. K. H. Smithe, A. J. Gabourie, V. Chen, S. Deshmukh, L. Bendersky, A. V. Davydov, E. Pop, "Ultrahigh thermal isolation across heterogeneously layered two-dimensional materials," *Sci. Adv.* vol. 5, no. 8, pp. eaax1325, 2019, doi: 10.1126/sciadv.aax1325.

[21] C. Lu, Y. Z. Sun, C. Wang, H. Zhang, W. Zhao, X. Hu, M. Xiao, W. Ding, Y. C. Liu, C. T. Chan, "On-chip nanophotonic topological rainbow," *Nature Commun.* vol. 13, no. 1, pp. 2586, 2022, doi: 10.1038/s41467-022-30276-w.

[22] Y. Baba, I. Matsuya, M. Nishikawa, T. Ishibashi, "Measurement of polarization properties of fifth harmonic signals in apertureless-type scanning near-field optical microscopy," *Japanese Journal of Appl. Phys.* vol. 57, no. 9S2, pp. 09TC04, 2018, doi: 10.7567/JJAP.57.09TC04.

[23] D. E. Tranca, S. G. Stanciu, R. Hristu, A. M. Ionescu, G. A. Stanciu, "Nanoscale local modification of PMMA refractive index by tip-enhanced femtosecond pulsed laser irradiation," *Appl. Sur. Sci.* vol. 623, pp. 157014, 2023, doi: 10.1016/j.apsusc.2023.157014.

[24] X. Yin, P. Shi, L. Du, X. Yuan, "Spin-resolved near-field scanning optical microscopy for mapping of the spin angular momentum distribution of focused beams," *Appl. Phys. Lett.* vol. 116, no. 24, pp. 241107, 2020, doi: 10.1063/5.0004750.

[25] P. Dey, "Fluorescence Microscope Confocal Microscope and Other Advanced Microscopes: Basic Principles and Applications in Pathology," In: *Basic and Advanced Laboratory Techniques in Histopathology and Cytology*, *Springer*, 2022, Singapore.

[26] M. Soltanmohammadi, E. Spurio, A. Gloystein, P. Luches, N. Nilius, "Photoluminescence Spectroscopy of Cuprous Oxide: Bulk Crystal versus Crystalline Films," *Phys. Status Solidi A* pp. 2200887, 2023, doi: 10.1002/pssa.202200887.

[27] D. Mrđenović, Z. F. Cai, Y. Pandey, G. L. Bartolomeo, R. Zenobi, N. Kumar, "Nanoscale chemical analysis of 2D molecular materials using tip-enhanced Raman spectroscopy," *Nanoscale* vol. 15, pp. 963-974, 2023, doi:10.1039/d2nr05127c.

[28] A. Dopilka, Y. Gu, J. M. Larson, V. Zorba, R. Kostecki, "Nano-FTIR Spectroscopy of the Solid Electrolyte Interphase Layer on a Thin-Film Silicon Li-Ion Anode," *ACS Appl. Mater. Interf.* vol. 15, no. 5, pp. 6755-6767 doi 10.1021/acsami.2c19484.

[29] P. Pellegrino, A. P. Bramanti, I. Farella, M. Cascione, V. De Matteis, A. Della Torre, F. Quaranta, R. Rinaldi, "Pulse-atomic force lithography: A powerful nanofabrication technique to fabricate constant and varying-depth nanostructures," *Nanomaterials* vol. 12, no. 6, pp. 991, 2022, doi: 10.3390/nano12060991.

[30] S. M. Aghaei, N. Yasrebi, B. Rashidian, "Characterization of line nanopatterns on positive photoresist produced by scanning near-field optical microscope," *J. Nanomater.* vol. 936876, pp. 1-7, 2015, doi: 10.1155/2015/936876.

[31] A. Roszkiewicz, A. Jain, M. Teodorczyk, W. Nasalski, "Formation and characterization of hole nanopattern on photoresist layer by scanning near-field optical microscope," *Nanomaterials* vol. 9, no. 10, pp. 1452, 2019, doi: 10.3390/nano9101452.

[32] G. Liu, M. Hirtz, H. Fuchs, Z. Zheng, "Development of Dip-Pen Nanolithography (DPN) and its derivatives," *Small* vol. 15, no. 21, pp. 1900564, 2019, doi: 10.1002/smll.201900564.

[33] S. T. Howell, A. Grushina, F. Holzner, J. Brugger, "Thermal scanning probe lithography—A review," *Microsystems & nanoengineering* vol. 6, no. 1, pp. 21, 2020, doi: 10.1038/s41378-019-0124-8.

[34] M. Behzadirad, A. K. Rishinaramangalam, D. Feezell, T. Busani, C. Reuter, A. Reum, M. Holz, T. Gotszalk, S. Mechold, M. Hofmann, A. Ahmad, T. Ivanov, I. W. Rangelow, "Field emission scanning probe lithography with GaN nanowires on active cantilevers," *Journal of Vacuum Science & Technology B Nanotech Microelectr: Materials Processing Measurement and Phenomena* vol. 38, no. 3, pp. 032806, 2020, doi.org/10.1116/1.5137901.

[35] K. J. Park, J. H. Huh, D. W. Jung, J. S. Park, G. H. Choi, G. Lee, P. J. Yoo, H. G. Park, G. R. Yi, S. Lee, "Assembly of "3D" plasmonic clusters by "2D" AFM nanomanipulation of highly uniform and smooth gold nanospheres," *Sci. Rep.* vol. 7, no. 1, pp. 6045, 2017, doi: 10.1038/s41598-017-06456-w.

[36] L. Zhao, T. Dai, Z. Qiao, P. Sun, J. Hao, Y. Yang, "Application of artificial intelligence to wastewaer treatment: A bibliometric analysis and systematic review of technology economy management and wastewater use," *Proc. Safety Environ. Prot.* vol. 133, pp. 169-182, 2020, doi: 10.1016/j.psep.2019.11.014.

[37] A. Malviya, D. Jaspal, "Artificial intelligence as an upcoming technology in wastewater treatment: a comprehensive review," *Environm. Tech. Reviews* vol. 10, no. 1, pp. 177-187, 2021, doi: 10.1080/21622515.2021.1913242.

[38] A. Pregowska, M. Osial, W. Urbańska, "The Application of Artificial Intelligence in the Effective Battery Life Cycle in the Closed Circular Economy Model—A Perspective," *Recycling* vol. 7, pp. 81, 2022, doi: 10.3390/recycling7060081.

[39] P. Zhang, Z. Guo, S. Ullah, G. Melagraki, A. Afantitis, I. Lynch, "Nanotechnology and artificial intelligence to enable susianable and precision agriculture," *Nature Plants* vol. 7, pp. 864-876, 2021, doi: 10.1038/s41477-021-00946-6.

[40] M. Tanzifi, S. H. Hosseini, A. D. Kiadehi, M. Olazar, K. Karimipour, R. Rezaiemehr, I. Ali, "Artificial neural network optimization for methyl orange adsorption onto polyaniline nano-adsorbent: Kinetic isotherm and thermodynamic studies," *J. Mol. Liquids* vol. 244, pp. 189-200, 2017, doi: 10.1016/j.molliq.2017.08.122.

[41] F. Mahmoodi, P. Darvishi, B. Vaferi, "Prediction of coefficients of the Langmuir adsorption isotherm using various artificial intelligence (AI) techniques," *J. Iran Chem. Soc.* vol. 15, pp. 2747–2757, 2018, doi: 10.1007/s13738-018-1462-4.

[42] M. Huang, Z. Li, H. Zhu, "Recent advances of graphene and related materials in Artificial Intelligence," *Adv. Int. Systems* vol. 4, no. 10, pp. 2200077, 2022, doi: 10.1002/aisy.202200077.

[43] N. X. Ho, T. T. Le, M. V. Le, "Development of artificial intelligence based model for the prediction of Young's modulus of polymer/carbon-nanotubes composites," *Mech. Adv. Mat. Structures* vol. 29, no. 27, pp. 5965-5978, 2022, doi: 10.1080/15376494.2021.1969709.

[44] M. Osial, A. Pregowska, "The Application of Artificial Intelligence in Magnetic Hyperthermia Based Research," *Future Internet* vol. 14, pp. 356, 2022, doi: 10.3390/fi14120356.

[45] B. Govindan, M. A. Sabri, A. Hai, F. Banat, M. A. Haija, "A review of Advanced Multifunctional Magnetic Nanostructures for Cancer Diagnosis and Therapy Integrated into an Artificial Intelligence Approach," *Pharmaceutics* vol. 15, pp. 868, 2023, doi: 10.3390/pharmaceutics15030868.

[46] E. Nsugbe, "An artificial intelligence-based decision support system for early diagnosis of polycystic ovaries syndrome," *Healthcare Anal.* vol. 3, pp. 100164, 2023, doi: 10.1016/j.health.2023.100164.

[47] O. Elemento, C. Leslie, J. Lundin, G. Tourassi, "Artificial intelligence in cancer research diagnosis and therapy," *Nat. Rev. Cancer* vol. 21, pp. 747–752, 2021, doi: 10.1038/s41568-021-00399-1.

[48] X. Chen, S. Xu, S. Shabani, Y. Zhao, M. Fu, A. J. Millis, M. M. Fogler, A. N. Pasupathy, M. Liu, D. N. Basov "Machine Learning for Optical Scanning Probe Nanoscopy," *Adv. Mater.* vol. 2109171, 2022, doi: 10.1002/adma.202109171.

[49] A. Konečná, F. Iyikanat, F. J. García de Abajo, "Entangling free electrons and optical excitations," *Sci. Adv.* vol. 8, no. 47, pp. eabo7853, 2022, doi: 10.1126/sciadv.abo7853.

[50] M. Ziatdinov, D. Kim, S. Neumayer, K. Rama, L. Vasudevan Collins, S. Jesse, M. Ahmadi, S. V. Kalinin, "Imaging mechanism for hyperspectral scanning probe microscopy via Gaussian process modelling," *Njp Comput. Mater.* vol. 6, pp. 21, 2020, doi:10.48550/arXiv.1911.11348.

[51] P. Fan, J. Gao, H. Mao, Y. Geng, Y. Yan, Y. Wang, S. Goel, X. Luo, "Scanning Probe Lithography: State-of-the-Art and Future Perspectives," *Micromachines* vol. 13, pp. 228, 2022, doi: 10.3390/mi13020228.

[52] M. Krenn, J. Landgraf, T. Foesel, F. Marquardt, "Artificial intelligence and machine learning for quantum technologies," *Phys. Rev. A* vol. 107, pp. 010101, 2023, doi: 10.1103/PhysRevA.107.010101.

[53] X. Chen, S. Xu, S. Shabani, Y. Zhao, M. Fu, A. J. Millis, M. M. Fogler, A. N. Pasupathy, M. Liu, D. N. Basov, "Machine Learning for Optical Scanning Probe Nanoscopy," *Adv. Mater.* vol. 35, pp. 2109171, 2023, doi: 101002/adma(2021)09171.

[54] A. Liberati, D. G. Altman, J. Tetzlaff, C. Mulrow, P. C. Gotzsche, J. P. A. Ioannidis, M. Clarke, P. J. Devereaux, J. Kleijnen, D. Moher, "The PRISMA Statement for Reporting Systematic Reviews and Meta-Analyses of Studies That Evaluate Health Care Interventions: Explanation and Elaboration," *PLoS Med.* vol. 6, pp. 1000100, 2009, doi: 10.1371/journal.pmed.1000100.

[55] M. L. Rethlefsen, S. Kirtley, S. Waffenschmidt, A. P. Ayala, D. Moher, M. J. Page, J. Koffel, "B PRISMA-S: an ex-tension to the PRISMA Statement for Reporting Literature Searches in Systematic Reviews," *Syst. Rev.* vol. 10, pp. 39, 2021.

[56] D. Gough, S. Oliver, J. Thomas, "An introduction to systematic reviews" (2nd ed) Sage, 2017.

[57] G. Binnig, H. Rohrer, C. Gerber, E. Weibel, "Tunneling through a controllable vacuum gap," *Appl. Phys. Lett.* vol. 40, pp. 178-180, 1982, doi: 10.1063/1.92999.

[58] D. W. Pohl, W. Denk, M. Lanz, "Optical stethoscopy: Image recording with resolution $\lambda/20$," *Appl. Phys. Lett.* vol. 44, pp. 651-653, 1984, doi: 10.1063/1.94865.

[59] A. Lewis, M. Isaacson, A. Harootunian, A. Muray, "Development of a 500 Å spatial resolution light microscope: I light is efficiently transmitted through $\lambda/16$ diameter apertures," *Ultramicroscopy* vol. 13, pp. 227-231, 1984, doi: 10.1016/0304-3991(84)90201-8.

[60] E. H. A. Synge, "A suggested method for extending microscopic resolution into the ultra-microscopic region," *The London Edinburgh and Dublin Philosophical Magazine and Journal of Science*. Series 7 vol. 6, no. 35, pp. 356–362, 2009, doi: 10.1080/14786440808564615.

[61] J. R. Matey, J. Blanc, "Scanning capacitance microscopy," *J. Appl. Phys.* vol. 57, pp. 1437-1444, 1985, doi: 10.1063/1.334506.

[62] G. Binnig, C. F. Quate, C. Gerber, "Atomic force microscope," *Phys. Rev. Lett.* vol. 56, pp. 930, 1986, doi: 10.1103/physrevlett.56.930.

[63] C. C. Williams, H. K. Wickramasinghe, "Scanning thermal profiler," *Microelectron Eng.* vol. 5, pp. 509-513, 1986, doi: 10.1016/0167-9317(86)90084-5.

[64] Y. Martin, H. K. Wickramasinghe, "Magnetic imaging by ''force microscopy'' with 1000 Å resolution," *Appl. Phys. Lett.* vol. 50, pp. 1455-1457, 1987, doi: 10.1063/1.97800.

[65] D. P. E. Smith, M. D. Kirk, C. F. Quate, "Molecular images and vibrational spectroscopy of sorbic acid with the scanning tunneling microscope," *J. Chem. Phys.* vol. 86, pp. 6034-6038, 1987, doi: 10.1063/1.452491.

[66] Y. Martin, D. W. Abraham, H. K. Wickramasinghe, "High-resolution capacitance measurement and potentiometry by force microscopy," *Appl. Phys. Lett.* vol. 52, pp. 1103-1105, 1988, doi: 10.1063/1.99224.

[67] J. H. Coombs, J. K. Gimzewski, B. Reihl, J. K. Sass, R. R. Schlittler, "Photon emission experiments with the scanning tunnelling microscope," *J. Microsc.* vol. 152, pp. 325-336, 1988, doi: 10.1111/j.1365-2818.1988.tb01393.x.

[68] O. E. Hüsser, D. H. Craston, A. J. Bard, "Scanning Electrochemical Microscopy: High-Resolution Deposition and Etching of Metals," *J. Electrochem. Soc.* vol. 136, pp. 3222, 1989, doi: 10.1149/1.2096429.

[69] P. K. Hansma, B. Drake, O. Marti, S. A. C. Gould, C. B. Prater, "The scanning ion-conductance microscope," *Science* vol. 243, pp. 641-643, 1989, doi: 10.1126/science.2464851.

[70] Y. Manassen, R. J. Hamers, J. E. Demuth, A. J. Castellano Jr, "Direct observation of the precession of individual paramagnetic spins on oxidized silicon surfaces," *Phys. Rev. Lett.* vol. 62, pp. 2531-2534, 1989, doi: 10.1103/PhysRevLett.62.2531.

[71] R. J. Hamers, K. Markert, "Surface photovoltage on Si(111)-(7×7) probed by optically pumped scanning tunneling microscopy," *J. Vac. Sci.* vol. 8, pp. 3524–3530, 1990, doi: 10.1116/1.576501.

[72] R. C. Reddick, R. J. Warmack, D. W. Chilcott, S. L. Sharp, T. L. Ferrell, "Photon scanning tunneling microscopy," *Rev. Sci. Instrum.* vol. 61, pp. 3669-3677, 1990, doi: 10.1063%2F1.1141534.

[73] C. C. Williams, H. K. Wickramasinghe, "Scanning chemical potential microscope: A new technique for atomic scale surface investigation," *J. Vac. Sci. Technol. B* vol. 9, pp. 537-540.

[74] M. Nonnenmacher, M. P. O'Boyle, H. K. Wickramasinghe, "Kelvin probe force microscopy," *Appl. Phys. Lett.* vol. 58, pp. 2921-2923,1991, doi: 10.1063/1.105227.

[75] F. Zenhausern, M. P. O'Boyle, H. K. Wickramasinghe, "Apertureless near-field optical microscope," *Appl. Phys. Lett.* vol. 65, pp. 1623-1625, 1994, doi: 10.1063/1.112931.

[76] Y. Seo, W. Jhe, "Atomic force microscopy and spectroscopy," *Rep. Prog. Phys.* vol. 71, pp. 0161011-23, doi: 10.1088/0034-4885/71/1/016101.

[77] N. Ishida, V. S. Craig, "Direct measurement of interaction forces between surfaces in liquids using atomic force microcopy," *Kona* vol. 36, pp. 187-200, 2019, doi: 10.14356/kona.2019013.

[78] R. Bowen, N. Hilal, "Atomic force microscopy in process engineering: An introduction to AFM for improved processes and products", Butterworth-Heinemann Oxford United Kingdom 2009

[79] Y. Zhang, K. Cui, Q. Gao, S. Hussain, Y. Lv, "Investigation of morphology and texture properties of WSi2 coatings on W substrate based on contact-mode AFM and EBSD," *Surf. Coat. Technol.* vol. 396, no. 125966, pp. 1-11, 2020, doi: 10.1016/j.surfcoat.2020.125966.

[80] A. C. V. Dos Santos, D. Tranchida, B. Lendl, G. Ramer, "Nanoscale chemical characterization of a post-consumer recycled polyolefin blend using tapping mode AFM-IR," *Analyst* vol. 147, pp. 3741-3747, 2022, doi: 10.1039/D2AN00823H.

[81] S. Freund, A. Hinaut, N. Marinakis, E. C. Constable, E. Meyer, C. E. Housecroft, T. Glatzel, "Anchoring of a dye precursor on NiO (001) studied by non-contact atomic force microscopy," *Beilstein J. Nanotechnol.* vol. 9, pp. 242-249, doi: org/10.3762/bjnano.9.26.

[82] F. P. Quacquarelli, J. Puebla, T. Scheler, D. Andres, C. Bödefeld, B. Sipos, C. D. Savio, A. Bauer, C. Pfleiderer, K. Karrai, "Scanning probe microscopy in an ultra-low vibration closed-cycle cryostat: Skyrmion lattice detection and tuning fork implementation," *Microscopy Today* vol. 23, pp. 12-17, 2015, doi: 10.1017/S1551929515000954.

[83] C. Barron, S. O'Toole, D. Zerulla, "Fabrication of Nanoscale Active Plasmonic Elements Using Atomic Force Microscope Tip-Based Nanomachining," *Nanomanuf* vol. 5, pp. 50-59, 2022, doi: 10.1007/s41871-021-00121-7.

[84] P. Lindner, L. Bargsten, S. Kovarik, J. Friedlein, J. Harm, S. Krause, R. Wiesendanger, "Temperature and magnetic field dependent behavior of atomic-scale skyrmions in Pd/Fe/Ir (111) nanoislands," *Phys. Rev. B* vol. 101, no. 214445, pp. 1-6, 2020, doi: 10.1103/PhysRevB.101.214445.

[85] C. Gusenbauer, E. Cabane, N. Gierlinger, J. Colson, J. Konnerth, "Visualization of the stimuli-responsive surface behavior of functionalized wood material by chemical force microscopy," *Sci. Rep.* vol. 9, pp. 1-9, 2019, doi: 10.1038/s41598-019-54664-3.

[86] K. L. Firestein, J. E. von Treifeldt, D. G. Kvashnin, J. F. Fernando, C. Zhang, A. G. Kvashnin, E. V. Podryabinkin, A. V. Shapeev, D. P. Siriwardena, P. B. Sorokin, D. Golberg, "Young's modulus and tensile strength of Ti3C2 MXene nanosheets as revealed by in situ TEM probing AFM nanomechanical mapping and theoretical calculations," *Nano Lett.* vol. 20, pp. 5900-5908, 2020, doi: 10.1021/acs.nanolett.0c01861.

[87] F. Lavini, F. Cellini, M. Rejhon, J. Kunc, C. Berger, W. de Heer, E. Riedo, "Atomic force microscopy phase imaging of epitaxial graphene films," *J. Phys. Materials* vol. 3, no. 024005, pp. 1-9, 2020, doi: 10.1088/2515-7639/ab7a02.

[88] M. Moreno-Moreno, P. Ares, C. Moreno, F. Zamora, C. Gomez-Navarro, J. Gomez-Herrero, "AFM manipulation of gold nanowires to build electrical circuits," *Nano Lett.* vol. 19, pp. 5459-5468, 2019, doi: 10.1021/acs.nanolett.9b01972.

[89] Y. Pandey, N. Kumar, G. Goubert, R. Zenobi, Nanoscale Chemical Imaging of Supported Lipid Monolayers using Tip-Enhanced Raman Spectroscopy," *Angew. Chem. Int. Ed.* vol. 60, pp. 19041-19046, 2021, doi: 10.1002/anie.202106128.

[90] C. K. Madawala, H. D. Lee, C. P. Kaluarachchi, A. V. Tivanski, "Probing the water uptake and phase state of individual sucrose nanoparticles using atomic force microscopy," *ACS Earth Space Chem.* vol. 5, pp. 2612-2620, 2021, doi: 10.1021/acsearthspacechem.1c00101.

[91] Q. Liu, H. Yang, "Application of atomic force microscopy in food microorganisms," *Trends Food Sci. Technol.* vol. 87, pp. 73-83, 2019, doi: 10.1016/j.tifs.2018.05.010.

[92] N. Piergies, E. Pięta, C. Paluszkiewicz, H. Domin, W. M. Kwiatek, "Polarization effect in tip-enhanced infrared nanospectroscopy studies of the selective Y5 receptor antagonist Lu AA33810," *Nano Res.* vol. 11, pp. 4401-4411, 2018, doi: 10.1007/s12274-018-2030-z.

[93] D. R. Cavalcanti DR, L. P. Silva, "Application of atomic force microscopy in the analysis of time since deposition (TSD) of red blood cells in bloodstains: A forensic analysis," *Forensic Sci. Int.* vol. 301, pp. 254-262, 2019, doi: 10.1016/j.forsciint.2019.05.048.

[94] A. Summerfield, M. Baldoni, D. V. Kondratuk, H. L. Anderson, S. Whitelam, J. P. Garrahan, E. Besley, P. H. Beton, "Ordering flexibility and frustration in arrays of porphyrin nanorings," *Nature Commun.* vol. 10, no. 1, pp. 2932, 2019, doi.org/10.1038/s41467-019-11009-y.

[95] J. A. Elemans, "Externally Applied Manipulation of Molecular Assemblies at Solid-Liquid Interfaces Revealed by Scanning Tunneling Microscopy," *Adv. Funct. Mat.* vol. 26, no. 48, pp. 8932-8951, doi: 10.1002/adfm.201603145.

[96] T. I. Zubar, V. M. Fedosyuk, S. V. Trukhanov, D. I. Tishkevich, D. Michels, D. Lyakhov, A. V. Trukhanov, "Method of surface energy investigation by lateral AFM: Application to control growth mechanism of nanostructured NiFe films," *Sci. Rep.* vol. 10, no. 1, pp. 1-10, 2020, doi: 10.1038/s41598-020-71416-w.

[97] S. Kim, D. Moon, B. R. Jeon, J. Yeon, X. Li, S. Kim, "Accurate Atomic-Scale Imaging of Two-Dimensional Lattices Using Atomic Force Microscopy in Ambient Conditions," *Nanomat.* vol. 12, no. 9, pp. 1542, 2022, doi: 10.3390/nano12091542.

[98] G. Pinto, P. Canepa, C. Canale, M. Canepa, O. Cavalleri, "Morphological and mechanical characterization of DNA SAMs combining nanolithography with AFM and optical methods," Materials vol. 13, no. 13, pp. 2888, 2020, doi: 10.3390/ma13132888.

[99] X. Shi, W. Qing, T. Marhaba, W. Zhang, "Atomic force microscopy-Scanning electrochemical microscopy (AFM-SECM) for nanoscale topographical and electrochemical characterization: Principles applications and perspectives," *Electrochimica Acta* vol. 332, pp. 135472, 2020, doi: 10.1016/j.electacta.2019.135472.

[100] B. Voigtländer, "Atomic Force Microscopy", 2nd ed. pp 137–147 Springer Berlin Germany, 2019.

[101] K. Muzyka, F. Rico, G. Xu, I. Casuso, "DNA at conductive interfaces: What can atomic force microscopy offer?," *J. Electroanal. Chem.* vol. 938, no. 117448, pp. 1-14, 2023, doi: 10.1016/j.jelechem.2023.117448.

[102] F. Sumbul, F. Rico, "Single-Molecule Force Spectroscopy: Experiments Analysis and Simulations In Atomic Force Microscopy: Methods and Protocols", N. C. Santos, F. A. Carvalho Eds, Springer New York 163-189, 2019.

[103] T. Wang, L. Jia, Q. Zhang, Z. Xu, Z. Huang, P. Yuan, B. Hou, X. Song, K. Nie, C. Liu, J. Wang, H. Yang, L. Liu, T. Zhang, Y. Wang, "Fabrication and Characterization of Pre-Defined Few-Layer Graphene," *Physchem.* vol. 3, pp. 13-21, 2023, doi: 10.3390/physchem3010002.

[104] J. Sanderson, "Multi-photon microscopy", *Current Protocols* vol. 3, pp. e634, 2023, doi: 10.1002/cpz1.634.

[105] T. Nörenberg, L. Wehmeier, D. Lang, S. C. Kehr, L. M. Eng, "Compensating for artifacts in scanning near-field optical microscopy due to electrostatics," *APL Photonics* vol. 6, no. 3, pp. 036102, 2021, doi: 10.1063/5.0031395.

[106] E. Sheremet, L. Kim, D. Stepanichsheva, V. Kolchuzhin, A. Milekhin, D. R. T. Zahn, R. D. Rodriguez, "Localized surface curvature artifacts in tip-enhanced nanospectroscopy imaging," *Ultramicroscopy* vol. 206, pp. 112811, 2019, doi: 10.1016/j.ultramic.2019.112811.

[107] B. Hecht, H. Bielefeldt, Y. Inouye, D. W. Pohl, L. Novotny, "Facts and artifacts in near-field optical microscopy," *J. Appl. Phys.* vol. 81, no. 6, pp. 2492-2498, 1997, doi: 10.1063/1.363956.

[108] J. H. Kindt, G. E. Fantner, J. A. Cutroni, P. K. Hansma, "Rigid design of fast scanning probe microscopes using finite element analysis," *Ultramicroscopy,* vol. 100, pp. 259-265, 2004, doi: 10.1016/j.ultramic.2003.11.009.

[109] R. Z. Liu, Z. Z. Shen, R. Wen, L. J. Wan, "Recent advances in the application of scanning probe microscopy in interfacial electroanalytical chemistry," *J. Electroanal. Chem.* vol. 938, pp. 117443, 2004, doi: 10.1016/j.jelechem.2023.117443.

[110] I. Sychugov, H. Omi, T. Murashita, Y. Kobayashi, "Modeling tip performance for combined STM-luminescence and aperture-SNOM scanning probe: Spatial resolution and collection efficiency," *Appl. Surf. Sci.* vol. 254, no. 23, pp. 7861-7863, 2008 doi: 10.1016/j.apsusc.2008.03.005.

[111] D. Becerril, T. Cesca, G. Mattei, C. Noguez, G. Pirruccio, M. Luce, A. Cricenti, "Active stabilization of a pseudoheterodyne scattering scanning near field optical microscope," *Rev. Sci. Instrum.* vol. 94, no. 2, pp. 023704, 2023, doi: 10.1063/5.0133488.

[112] C. Hafner, R. Hiptmair, P. Souzangar, "Data-sparse numerical models for SNOM tips," *Int. J. Numer. Model.* vol. 30, pp. e2178, 2017, doi: 10.1002/jnm.2178.

[113] Q. Qian, H. Yu, P. Gou, J. Xu, Z. An, "Plasmonic focusing of infrared SNOM tip patterned with asymmetric structures," *Opt. Express* vol. 23, no. 10, pp. 12923-12934, 2015, doi: 10.1364/OE.23.012923.

[114] X. Wang, J. Cui, H. Yin, Z. Wang, X. He, X. Mei, "Mechanism of near-field optical nanopatterning on noble metal nano-films by a nanosecond laser irradiating a cantilevered scanning near-field optical microscopy probe," *Appl. Opt.* vol. 62, pp. 3672-3682, 2023, doi: 10.1364/AO.487295.

[115] X. Guo, C. Wu, S. Zhang, D. Hu, S. Zhang, Q. Jiang, X. Dai, Y. Duan, X. Yang, Z. Sun, S. Zhang, H. Xu, Q. Dai, "Mid-infrared analogue polaritonic reversed Cherenkov radiation in natural anisotropic crystals," *Nat. Commun.* vol. 14, pp. 2532, 2023, doi: 10.1038/s41467-023-37923-w.

[116] G. Hu, W. Ma, D. Hu, J. Wu, C. Zheng, K. Liu, X. Zhang, X. Ni, J. Chen, X. Zhang, Q. Dai, J. D. Caldwell, A. Paarmann, A. Alù, P. Li, C. W. Qiu, "Real-space nanoimaging of hyperbolic shear polaritons in a monoclinic crystal," *Nat. Nanotechnol.* vol. 18, pp. 64–70, 2023, doi: 10.1038/s41565-022-01264-4.

[117] G. Lu, Z. Pan, C. R. Gubbin, R. A. Kowalski, S. De Liberato, D. Li, J. D. Caldwell, "Launching and Manipulation of Higher-Order In-Plane Hyperbolic Phonon Polaritons in Low-Dimensional Heterostructures," *Adv. Mater.* vol. 35, pp. 2300301, 2023, doi: 10.1002/adma.202300301.

[118] X. Guo, N. Li, X. Yang, R. Qi, C. Wu, R. Shi, Y. Li, Y. Huang, F. J. García de Abajo, E. G. Wang, P. Gao, Q. Dai, "Hyperbolic whispering-gallery phonon polaritons in boron nitride nanotubes," *Nat. Nanotechnol. vol.* 18, pp. 529–534, 2023, doi: 10.1038/s41565-023-01324-3.

[119] N. Granchi, M. Lodde, K. Stokkereit, R. Spalding, P. J. van Veldhoven, R. Sapienza, A. Fiore, M. Gurioli, M. Florescu, F. Intonti, "Near-field imaging of optical nanocavities in hyperuniform disordered materials," *Phys. Rev. B* vol. 107, no. 6, pp. 064204, 2023, doi: 10.1103/PhysRevB.107.064204.

[120] P. Lalanne, W. Yan, K. Vynck, C. Sauvan, J. P. Hugonin, "Light interaction with photonic and plasmonic resonances," *Laser Photonics Rev.* vol. 12, pp. 1, 2018, doi: 10.1002/lpor.201700113.

[121] Z. Fei, A. S. Rodin, G. O. Andreev, W. Bao, A. S. McLeod, M. Wagner, L. M. Zhang, Z. Zhao, M. Thiemens, G. Dominguez, M. M. Fogler, A. H. Castro Neto, C. N. Lau, F. Keilmann, D. N. Basov, "Gate-tuning of graphene plasmons revealed by infrared nano-imaging," *Nature* vol. 487, pp. 82, 2012, doi: 10.1038/nature11253.

[122] Y. Luan, M. Kolmer, M. C. Tringides, Z. B. Fei, "Nanoscale infrared imaging and spectroscopy of hot-electron plasmons in graphene," *Phys. Rev.* vol. 107, no. 8, pp. 085414, 2023, doi: 10.1103/PhysRevB.107.085414.

[123] R. K. Jain, V. K. Jain, P. M. Dixit, "Modeling of material removal and surface roughness in abrasive flow machining process," *Int. J. Mach. Tools Manuf.* vol. 39, no. 12, pp. 1903–1923, 1999, doi: 10.1016/S0890-6955(99)00038-3.

[124] X. Liang, T. Kojima, M. Ito, N. Amino, H. Liu, M. Koishi, K. Nakajima, "In Situ (2023) Nanostress Visualization Method to Reveal the Micromechanical Mechanism of Nanocomposites by Atomic Force Microscopy," *ACS Appl. Mater. Interfaces* vol. 15, no. 9, pp. 12414-12422, 1999, doi: 10.1021/acsami.2c22971.

[125] B. Xue, E. Brousseau, C. Bowen, "Modelling of a shear-type piezoelectric actuator for AFM-based vibration-assisted nanomachining," *Int. J. Mech. Sci.* vol. 243, pp. 108048, 2023, doi: 10.1016/j.ijmecsci.2022.108048.

[126] O. M. Primera-Pedrozo, S. Tan, D. Zhang, B. T. O'Callahan, W. Cao, E. T. Baxter, X. B. Wang, P. Z. El-Khoury, V. Prabhakaran, V. A. Glezakou, G. E. Johnson, "Influence of surface and intermolecular interactions on the properties of supported polyoxometalates," *Nanoscale* vol. 15, pp. 5786-5797, 2023, doi: 10.1039/D2NR06148A.

[127] J. Schirmer, R. Chevigny, A. Emelianov, E. Hulkko, A. Johansson, P. Myllyperkiö, E. D. Sitsanidis, M. Nissinen, M. Pettersson, "Diversity at the nanoscale: laser-oxidation of single-layer graphene affects Fmoc-phenylalanine surface-mediated self-assembly," *Phys. Chem. Chem. Phys.* Vol. 2, pp. 8725-8733, 2023, doi: 10.1039/D3CP00117B.

[128] D. V. Grudinin, G. A. Ermolaev, D. G. Baranov, A. N. Toksumakov, K. V. Voronin, A. S. Slavich, A. A. Vyshnevyy, A. B. Mazitov, I. A. Kruglov, D. A. Ghazaryan, A. V. Arsenin, K. S. Novoselov, V. S. Volkov, "Hexagonal boron nitride nanophotonics: a record-breaking material for the ultraviolet and visible spectral ranges," *Mater. Horiz.* Vol. 10, pp. 2427-2435, 2023, doi: 10.1039/D3MH00215B.

[129] W. Xiong, J. Lu, J. Geng, Z. Ruan, H. Zhang, Y. Zhang, G. Niu, B. Fu, Y. Zhang, S. Sun, L. Gao, J. Cai, "Atomic-scale construction and characterization of quantum dots array and poly-fluorene chains via 27-dibromofluorene on Au(111)," *Appl. Surf. Sci.* vol. 609, pp. 155315, 2023, doi: 10.1016/j.apsusc.2022.155315.

[130] B. Xing, R. Gao, M. Wu, H. Wei, S. Chi, Z. Hua, "Differentiation on crystallographic orientation dependence of hydrogen diffusion in α-Fe and γ-Fe: DFT calculation combined with SKPFM analysis," *Appl. Surf. Sci.* vol. 615, pp. 156395, 2023, doi: 10.1016/j.apsusc.2023.156395.

[131] A. A. Govyadinov, S. Mastel, F. Golmar, A. Chuvilin, P. S. Carney, R. Hillenbrand, "Recovery of Permittivity and Depth from Near-Field Data as a Step Towards Optical Nanotomography," *ACS Nano* vol. 8, no. 7, pp. 6911, 2014, doi: 10.1021/nn5016314.

[132] X. Guo, X. He, Z. Degnan, C. C. Chiu, B. C. Donose, K. Bertling, A. Fedorov, A. D. Rakic, P. Jacobson, "Terahertz nanospectroscopy of plasmon polaritons for the evaluation of doping in quantum devices," *Nanophotonics* vol. 12, no. 10, pp. 1865-1875, 2023, doi: 10.1515/nanoph-2023-0064.

[133] R. Hillenbrand, B. Knoll, F. Keilmann, "Pure optical contrast in scattering-type scanning near-field microscopy," *J. Microsc.* Vol. 202, pp. 77–83, 2001, doi: 10.1046/j.1365-2818.2001.00794.x.

[134] D. Siebenkotten, B. Kaestner, A. Hoehl, S. Amakawa, "Calibration method for complex permittivity measurements using s-SNOM combining multiple tapping harmonics", 2023, arXiv preprint arXiv:230517031.

[135] M. Azib, F. Baudoin, N. Binaud, C. Villeneuve-Faure, F. Bugarin, S. Segonds, G. Teyssedre, "Numerical simulations for quantitative analysis of electrostatic interaction between atomic force microscopy probe and an embedded electrode within a thin dielectric: meshing optimization sensitivity to potential distribution and impact of cantilever contribution," *J. Phys. D: Appl. Phys.* vol. 51, no. 16, pp. 165302, 2018, doi: 10.1088/1361-6463/aab286.

[136] Z. Dong, Z. Liu, P. Wang, X. Gong, "Nanostructure characterization of asphalt-aggregate interface through molecular dynamics simulation and atomic force microscopy," *Fuel* vol. 189, no. 155-163, 2017, doi: 10.1016/j.fuel.2016.10.077.

[137] I. Argatov, X. Jin, G. Mishuris, "Atomic force microscopy-based indentation of cells: modelling the effect of a pericellular coat," *J. R. Soc. Interface* vol. 20, no. 199, pp. 20220857, 2023, doi: 10.1098/rsif.2022.0857.

[138] C. Valero, B. Navarro, D. Navajas, J. M. García-Aznar, "Finite element simulation for the mechanical characterization of soft biological materials by atomic force microscopy," *J. Mech. Behav. Biomed. Mater.* vol. 62, pp. 222-235, 2016, doi: 10.1016/j.jmbbm.2016.05.006.

[139] A. Gaveau, C. Coetsier, C. Roques, P. Bacchin, E. Dague, C. Causserand, "Bacteria transfer by deformation through microfiltration membrane," *J. Membr. Sci.* vol. 523, pp. 446-455, 2017, doi: 10.1016/j.memsci.2016.10.023.

[140] Y. Liu, K. P. Kelley, R. K. Vasudevan, H. Funakubo, M. A. Ziatdinov, S. V. Kalinin, "Experimental discovery of structure–property relationships in ferroelectric materials via active learning," *Nat. Mach. Intell.* vol. 4, pp. 341–350, 2022.

[141] Y. Liu, A. N. Morozovska, E. A. Eliseev, K. P. Kelley, R. Vasudevan, M. Ziatdinov, S. V. Kalinin, "Autonomous scanning probe microscopy with hypothesis learning: Exploring the physics of domain switching in ferroelectric materials," *Patterns* vol. 4, no. 100704, pp. 1-10, 2023, doi: 10.1016%2Fj.patter.2023.100704.

[142] N. Dietler, M. Minder, V. Gligorovski, A. M. Economou, D. A. H. L. Joly, A. Sadeghi, C. H. M. Chan, M. Koziński, M. Weigert, A. F. Bitbol, S. J. Rahi, "A convolutional neural network segments yeast microscopy images with high accuracy," *Nat. Commun.* vol. 11, pp. 5723, 2020, doi: 10.1038/s41467-020-19557-4.

[143] J. Gao, Q. Jiang, B. Zhou, D. Chen, "Convolutional neural networks for computer-aided detection or diagnosis in medical image analysis: An overview," *Mathematical Biosciences and Engineering* vol. 16, no. 6, pp. 6536-6561, 2019, doi: 10.3934/mbe.2019326.

[144] R. K. Vasudevan, K. P. Kelley, J. Hinkle, H. Funakubo, S. Jesse, S. V. Kalinin, M. Ziatdinov, "Autonomous Experiments in Scanning Probe Microscopy and Spectroscopy: Choosing Where to Explore Polarization Dynamics in Ferroelectrics," *ACS Nano* vol. 15, pp. 11253-11262, 2021, doi: 10.1021/acsnano.0c10239.

[145] S. V. Kalinin, S. Ziatdinov, J. Hinkle, S. Jesse, A. Ghosh, K. P. Kelley, A. R. Lupini, B. G. Sumpter, R. K. Vasudevan, "Automated and Autonomous Experiments in Electron and Scanning Probe Microscopy," *ACS Nano* vol. 15, no. 8, pp. 12604-12627, 2021, doi: 10.1021/acsnano.1c02104.

[146] Y. LeCun, Y. Bengio, G. Hinton, "Deep learning" *Nature* vol. 521, pp. 436–444, 2015, doi: 10.1038/nature14539.

[147] I. Goodfellow, Y. Bengio, A. Courville, "A Deep learning", Cambridge MA MIT Press, 2016.

[148] I. Malkiel, M. Mrejen, A. Nagler, U. Arieli, L. Wolf, H. Suchowski, "Plasmonic nanostructure design and characterization via Deep Learning Light," *Sci. Appl.* vol. 7, pp. 60, 2018, doi: 10.1038/s41377-018-0060-7.

[149] K. Yao, R. Unni, Y. Zheng, "Intelligent nanophotonics: merging photonics and artificial intelligence at the nanoscale," *Nanophotonics* vol. 8, no. 3, pp. 339-366, 2019, doi: 10.1515/nanoph-2018-0183.

[150] Z. Liu, D. Zhu, S. P. Rodrigues, K. T. Lee, W. Cai, "Generative Model for the Inverse Design of Metasurfaces," *Nano Letters* vol. 18, no. 10, pp. 6570-6576, 2018, doi: 10.1021/acs.nanolett.8b03171.

[151] T. Pu, J. Y. Ou, N. Papasimakis, N. I. Zheludev, "Label-free deeply subwavelength optical microscopy," *Appl. Phys. Lett.* vol. 116, no. 13, pp. 131105, 2020, doi.org/10.1063/5.0003330.

[152] B. Huang, Z. Li, J. Li, "An artificial intelligence atomic force microscope enabled by machine learning," *Nanoscale* vol. 10, pp. 21320-21326, 2018, doi: 10.1039/C8NR06734A.

[153] A. Krull, P. Hirsch, C. Rother, A. Schiffrin, C. Krull, "Artificial-intelligence-driven scanning probe microscopy," *Commun. Phys.* vol. 3, pp. 54, 2000, doi: 10.1038/s42005-020-0317-3.

[154] B. Alldritt, P. Hapala, N. Oinonen, F. Urtev, O. Krejci, F. F. Canova, J. Kannala, F. Schulz, P. Liljeroth, A. S. Foster, "Automated structure discovery in atomic force microscopy," *Sci. Adv.* vol. 26, no. 6, pp. 9:eaay6913, 2020, doi: 10.1126/sciadv.aay6913.

[155] A. Krizhevsky, I. Sutskever, G. E. Hinton, "Imagenet classification with deep convolutional neural networks," in: F. Pereira, C. J. C. Burges, L. Bottou, K. Q. Weinberger eds, Advances in Neural Information Processing System 25 Lake Tahoe NV USA NIPS 1097–105, 2012, doi: 10.1145/3065386.

[156] I. Azuri, I. Rosenhek-Goldian, N. Regev-Rudzki, G. Fantner, S. R. Cohen, "The role of convolutional neural networks in scanning probe microscopy: a review," *Beilstein J. Nanotechnol.* vol. 12, pp. 878–901, 2021, doi: 10.3762/bjnano.12.66.

[157] Q. Duan, Z. Xu, S. Zheng, J. Chen, Y. Feng, L. Run, J. Lee, "Machine learning based on holographic scattering spectrum for mixed pollutants analysis," *Anal. Chim. Acta* vol. 1143, pp. 298-305, 2021, doi: 10.1016/j.aca.2020.10.060.

[158] S. Xu, A. S. McLeod, X. C. Chen, R. J. Rizzo, B. S. Jessen, Z. Yao, Z. Wang, Z. Sun, S. Shabani, A. N. Pasupathy, A. J. Millis, C. R. Dean, J. C. Hone, M. Liu, D. N. Basov, "Deep Learning Analysis of Polaritonic Wave Images," *ACS Nano* vol. 15, no. 11, pp. 18182-18191, 2021, doi: 10.1021/acsnano.1c07011.

[159] M. H. Mozaffari, P. Abdolghader, L. L. Tay, A. Stolow, "Segmentation of Stimulated Raman Microscopy Images using a 1D Convolutional Neural Network," *Photonics North IEEE* vol. 1, 2022, doi: 10.1109/PN56061.2022.9908347.

[160] Y. Cao, V. Fatemi, A. Demir, S. Fang, S. L. Tomarken, J. Y. Luo, J. D. Sanchez-Yamagishi, K. Watanabe, T. Taniguchi, E. Kaxiras, R. C. Ashoori, P. Jarillo-Herrero, "Correlated insulator behaviour at half-filling in magic-angle graphene superlattices," *Nature* vol. 556, no. 7699, pp. 80-84, 2018, doi: 10.1038/nature26154.

[161] B. Peng, Q. Zhang, Y. Zhang, Y. Zhao, S. Hou, Y. Yang, F. Dai, R. Yi, R. Chen, J. Wang, L. Zhang, L. Chen, S. Zhang, H. Fang, "Unexpected Piezoresistive Effect Room-Temperature Ferromagnetism and Thermal Stability of 2D β-CuI Crystals in Reduced Graphene Oxide Membrane," *Adv. Electron. Mater.* vol. 2201241, 2023, doi: 10.1002/aelm.202201241.

[162] L. Chen, S. Li, Q. Bai, J. Yang, S. Jiang, Y. Miao, "Review of Image Classification Algorithms Based on Convolutional Neural Networks," *Remote Sens.* vol. 13, pp. 4712, 2021, doi: 10.3390/rs13224712.

[163] M. Rashidi, R. A. Wolkow, "Autonomous Scanning Probe Microscopy in Situ Tip Conditioning through Machine Learning," *ACS Nano* vol. 12, no. 6, pp. 5185–5189, 2018, doi: 10.1021/acsnano.8b02208.

[164] D. P. Kingma, J. Ba, "Adam: a method for stochastic optimization," 2014, Preprint at https://arxivorg/abs/14126980.

[165] B. Alldritt, F. Urtev, N. Oinonen, M. Aapro, J. Kannala, P. Liljeroth, A. S. Foster, "Automated tip functionalization via machine learning in scanning probe microscopy," *Computer Phys. Commun.* vol. 273, 2022, doi: 10.1016/j.cpc.2021.108258.

[166] L. Burzawa, S. Liu, E. W. Carlson, "Classifying surface probe images in strongly correlated electronic systems via machine learning," *Phys. Rev. Mater.* vol. 3, pp. 033805, 2019, doi: 10.1103/PhysRevMaterials.3.033805.

[167] E. W. Carlson, S. Liu, B. Phillabaum, K. A. Dahmen, "Decoding Spatial Complexity in Strongly Correlated Electronic Systems," *J. Supercond. Nov. Magn.* vol. 28, pp. 1237–1243, 2015, doi: 10.1007/s10948-014-2898-0.

[168] D. Menaka, S. G. Vaidyanathan, "A hybrid convolutional neural network-support vector machine architecture for classification of super-resolution enhanced chromosome images," *Expert Systems* vol. 40, no. 3, pp. e13186, 2023, doi: 10.1111/exsy.13186.

[169] W. S. Lai, J. B. Huang, N. Ahuja, M. H. Yang, "Fast and accurate image super-resolution with deep Laplacian pyramid networks," *IEEE Transactions on Pattern Analysis and Machine Intelligence* vol. 41, no. 11, pp. 2599–2613, 2019, doi: 10.1109/TPAMI.2018.2865304.

[170] C. H. D. Tsai, C. H. Yeh, "Neural Network for Enhancing Microscopic Resolution Based on Images from Scanning Electron Microscope," *Sensors* vol. 21, pp. 2139, 2021, doi: 10.3390/s21062139.

[171] M. Corrias, L. Papa, I. Sokolović, V. Birschitzky, A. Gorfer, M. Setvin, M. Schmid, U. Diebold, M. Reticcioli, C. Franchini, "Automated real-space lattice extraction for atomic force microscopy images," *Machine Learning: Science and Technology* vol. 4, no. 1, pp. 015015, 2023, doi: 10.1088/2632-2153/acb5e0.

[172] M. H. Mozaffari, L. L. Tay, "Independent Component Analysis for Spectral Unmixing of Raman Microscopic Images of Single Human Cells," in *Science and Information Conference Cham: Springer International Publishing* 204-2013, 2022, doi: 10.1007/978-3-031-10467-1_12.

[173] M. Ziatdinov, Y. Liu, A. N. Morozovska, E. A. Eliseev, X. Zhang, I. Takeuchi, S. V. Kalinin, "Hypothesis Learning in Automated Experiment: Application to Combinatorial Materials Libraries," *Adv. Mater.* vol. 34, pp. 2201345, 2022, doi: 10.1002/adma.202201345.

[174] R. Chi, H. Li, D. Shen, Z. Hou, B. Huang, "Enhanced P-type control: Indirect adaptive learning from set-point updates," *IEEE Transactions on Automatic Control* vol. 68, no. 3, pp. 1600-1613, 2022, doi: 10.1109/TAC.2022.3154347.

[175] B. G. Ellis, C. A. Whitley, S. A. Jedani, C. I. Smith, P. J. Gunning, P. Harrison, P. Unsworth, P. Gardner, R. J. Shaw, S. D. Barrett, A. Triantafyllou, J. M. Risk, P. Weightman, "Insight into metastatic oral cancer tissue from novel analyses using FTIR spectroscopy and aperture IR-SNOM," *Analyst* vol. 146, pp. 4895-4904, 2021, doi: 10.1039/D1AN00922B.

[176] G. Ciasca, A. Mazzini, T. E. Sassun, M. Nardini, E. Minelli, M. Papi, V. Palmieri, M. de Spirito, "Efficient Spatial Sampling for AFM-Based Cancer Diagnostics: A Comparison between Neural Networks and Conventional Data Analysis," *Condens. Matter* vol. 4, pp. 58, 2019, doi: 10.3390/condmat4020058.

[177] N. Borodinov, S. Neumayer, S. V. Kalinin, O. S. Ovchinnikova, R. K. Vasudevan, S. Jesse, "Deep neural networks for understanding noisy data applied to physical property extraction in scanning probe microscopy," *npj Computational Mater.* vol. 5, pp. 25, 2019, doi: 10.1038/s41524-019-0148-5.

[178] Barnard AS, Motevalli B, Parker A J, Fischer J M, Feigla C A, Opletal G (2019) Nanoinformatics, and the big challenges for the science of small things Nanoscale 11:19190-19201 https://doi.org/10.1039/C9NR05912A

[179] M. D. Wilkinson, M. Dumontier, I. J. Aalbersberg, G. Appleton, M. Axton, A. Baak, N. Blomberg, J. W. Boiten, L. B. da Silva Santos, P. E. Bourne, J. Bouwman, A. J. Brookes, T. Clark, M. Crosas, I. Dillo, O. Dumon, S. Edmunds, C. T. Evelo, R. Finkers, A. Gonzalez-Beltran, A. J. G. Gray, P. Groth, C. Goble, J. S. Grethe, J. Heringa, P. A. C. 't Hoen, R. Hooft, T. Kuhn, R. Kok, J. Kok, S. J. Lusher, M. E. Martone, A. Mons, A. L. Packer, R. Persson, P. Rocca-Serra, M. Roos, R. van Schaik, S. A. Sansone, E. Schultes, T. Sengstag, G. Strawn, M. A. Swertz, M. Thompson, J. van der Lei, E. van Mulligen, J. Velterop, A. Waagmeester, P. Wittenburg, K. Wolstencroft, J. Zhao, B. Mons, "The FAIR Guiding Principles for scientific data management and stewardship," *Sci. Data* vol. 3, pp. 160018, 2016, doi.org/10.1038/sdata.2016.18.

[180] T. Rodani, E. Osmenaj, O. Cazzaniga, M. Panighel, A. Cristina, S. Cozzini, "Towards the FAIRification of Scanning Tunneling Microscopy Images Data," *Intelligence* vol. 5, no. 1, pp. 27–42, 2023, doi: 10.1162/dint_a_00164.

[181] Z. Wu, R. Ramsundar, E. N. Feinberg, J. Gomes, C. Geniesse, A. S. Pappu, K. Leswingd, V. Pande, "MoleculeNet: a benchmark for molecular machine learning," *Chem. Sci.* vol. 9, pp. 513-530, 2018, doi: 10.1039/C7SC02664A.

[182] SPM Portal https://spmportalquasarsrcom/spmportal/user-guide (accesses on 2005(2023)).

[183] JARVIS (Joint Automated Repository for Various Integrated Simulations) https://jarvisnistgov/jarvisstm (accessed on 2005(2023)).

[184] K. Choudhary, K. F. Garrity, C. Camp, S. V. Kalinin, R. Vasudevan, M. Ziatdinov, F. Tavazza, "Computational scanning tunneling microscope image database," *Sci. Data* vol. 8, pp. 57, 2021, doi: 10.1038/s41597-021-00824-y.

[185] Y. Zhu, K. Yu, "Artificial intelligence (AI) for quantum and quantum for AI," *Opt. Quant. Electron.* vol. 55, pp. 697, 2023, doi: 101007/s11082-023-04914-6.

[186] V. Moret-Bonillo, "Can artificial intelligence benefit from quantum computing?," *Prog. Artif. Intell.* vol. 3, pp. 89–105, 2015, doi: 101007/s13748-014-0059-0.

[187] J. Zhang, Z. Chen, S. Mills, T. Ciavatti, Z. Yao, R. Mescall, H. Hu, Y. Semenenko, Z. Fei, H. Li, V. Perebeinos, H. Tao, Q. Dai, X. Du, M. Liu, "Terahertz nanoimaging of graphene," *ACS Photonics* vol. 5, no. 7, pp. 2645-2651, 2018, doi: 101021/acsphotonics8b00190.

[188] X. Guo, K. Bertling, A. D. Rakić, "Optical constants from scattering-type scanning near-field optical microscope," *Appl. Phys. Lett.* vol. 118, pp. 041103, 2021, doi: 101063/50036872.

[189] L. Luo, M. Mootz, J. H. Kang, C. Huang, K. Eom, J. W. Lee, C. Vaswani, Y. G. Collantes, E. E. Hellstorm, I. E. Perakis, C. B. Eom, J. Wang, "Quantum coherence tomography of light-controlled superconductivity," *Nat. Phys.* vol. 19, pp. 201–209, 2023, doi: 101038/s41567-022-01827-1.

[190] F. Flöther, "The state of quantum computing applications in health and medicine Research Directions," *Quantum Technologies* vol. 1 pp. E10, 2023, doi: 101017/qut(2023)4.

[191] L. Junyu, N. Khadijeh, S. Kunal, T. Francesco, J. Liang, M. Antonio, "Analytic Theory for the Dynamics of Wide Quantum Neural Networks," *Phys. Rev. Lett.* vol. 130, no. 15, pp. 150601, 2023, doi: 101103/PhysRevLett130150601.

[192] M. Larocca, N. Ju, D. García-Martín, P. J. Colez, M. Cerezo, "Theory of overparametrization in quantum neural networks," *Nat. Comput. Sci.* vol. 3, pp. 542–551, 2023, doi: 101038/s43588-023-00467-6.

[193] S. Sim, P. D. Johnson, A. Aspuru-Guzik, "Expressibility and entangling capability of parameterized quantum circuits for hybrid quantum-classical algorithms," *Adv. Quantum Technol.* vol. 2, no. 12, pp. 1900070, 2019, doi: 10.48550/arXiv.1905.10876.

[194] P. Gupta, L. S. Schadler, R. Sundararaman, "Dielectric properties of polymer nanocomposite interphases from electrostatic force microscopy using machine learning," *Materials Characterization* vol. 173, pp. 110909, 2021, doi: 10.1021/acsaelm.2c01331.

[195] R. Joseph, A. Farhadi, *YOLOv3: An Incremental Improvement Technical Report*, 2018.

[196] P. Gupta, E. Ruzicka, B. C. Benicewicz, R. Sundararaman, L. S. Schadler, "Dielectric Properties of Polymer Nanocomposite Interphases Using Electrostatic Force Microscopy and Machine Learning," *ACS Appl. Electron. Mater.* vol. 5, no. 2, pp. 794–802, 2023, doi: 10.1021/acsaelm.2c01331.

[197] H. Bai, S. Wu, "Deep-learning-based nanowire detection in AFM images for automated nanomanipulation," *Nanotech. Precision Engineering* vol. 4, pp. 013002, 2021, doi: 10.1063/10.0003218.

[198] S. Basak, M. A. Banguero, L. Burzawa, F. Simmons, P. Salev, L. Aigouy, M. M. Qazilbash, L. K. Schuller, D. K. Basov, A. Zimmers, E. W. Carlson, "Deep learning Hamiltonians from disordered image data in quantum materials," *Phys. Rev. B* vol. 107, pp. 205121, 2023, doi: 10.1103/PhysRevB.107.205121.

[199] open database https://githubcom/usnistgov/jarvis (access on December 1 (2023)).

[200] open database https://alex-krullgithubio/stm-datahtml (access on December 1 (2023)).

[201] BioImLab http://BioImLabdeiunipdit/Chromosome (access on December 1 (2023)).

[202] J. C. Caicedo, S. Cooper, F. Heigwer, S. Warchal, P. Qiu. C. Molnar, A. S. Vasilevich, J. D. Barry, H. S. Bansal, O. Kraus, M. Wawer, L. Paavolainen, M. D. Herrmann, M. Rohban, J. Hung, H. Hennig, J. Concannon, I. Smith, P. A. Clemons, S. Singh, P. Rees, P. Horvath, R. G. Linington, A. E. Carpenter, "Data-analysis strategies for image-based cell profiling," *Nature methods* vol. 14, no. 9, pp. 849-863, 2017, doi: 10.1038/nmeth.4397.

[203] O. Gordon, P. D'Hondt, L. Knijff, S. E. Freeney, F. Junqueira, P. Moriarty, I. Swart, "Scanning tunneling state recognition with multi-class neural network ensembles," *Review of Scientific Instruments* vol. 90, pp. 10, 2019, doi: 10.1063/1.5099590.

[204] O. Schoppe, C. Pan, J. Coronel, H. Mai, Z. Rong, M. I. Todorov, A. Müskes, F. Navarro, H. Li, A. Ertürk, B. H. Menze, "Deep learning-enabled multi-organ segmentation in whole-body mouse scans," *Nature communications* vol. 11, no. 1, pp. 5626, 2020, doi: 10.1038/s41467-020-19449-7.

[205] J. Sotres, H. Boyd, J. F. Gonzalez-Martinez, "Enabling autonomous scanning probe microscopy imaging of single molecules with deep learning," *Nanoscale* vol. 13, no. 20, pp. 9193-9203, 2021, doi: 10.1039/D1NR01109J.

[206] X. Li, L. Collins, K. Miyazawa, T. Fukuma, S. Jesse, S. V. Kalinin, "High-veracity functional imaging in scanning probe microscopy via Graph-Bootstrapping," *Nature Commun.* vol. 9, no. 1, pp. 2428, 2018, doi: 10.1038/s41467-018-04887-1.

[207] M. F. Liz, A. V. Nartova, A. V. Matveev, A. G. Okunev, "Using computer vision and deep learning for nanoparticle recognition on scanning probe microscopy images: modified U-net approach," *IEEE Science and Artificial Intelligence conference (SAI ence)* pp. 13-16, 2020, doi: 10.1109/S.A.I.ence50533.2020.9303184.

[208] Z. Li, K. Kamnitsas, B. Glocker, "Analyzing overfitting under class imbalance in neural networks for image segmentation," *IEEE transactions on medical imaging* vol. 40, no. 3, pp. 1065-1077, 2020, doi: 10.1109/TMI.2020.3046692.

[209] D. Enke, N. Mehdiyev, "A new hybrid approach for fo recasting interest rates," *Procedia Computer Science* vol. 12, pp. 259-264, 2012, doi: 10.1016/j.procs.2012.09.066.

[210] T. Hu, W. Wang, C. Lin, G. Cheng, "Regularization matters: A nonparametric perspective on overparametrized neural network," *PMLR International Conference on Artificial Intelligence and Statistics* pp. 829-837, 2021.

[211] J. Lüder, "Determining electronic properties from L-edge x-ray absorption spectra of transition metal compounds with artificial neural networks," *Phys. Rev. B* vol. 103, no. 4, pp. 045140, 2021, doi: 10.48550/arXiv.2009.09684.

[212] T. Rodani, E. Osmenaj, A. Cazzaniga, M. Panighel, A. Cristina, S. Cozzini, "Towards the FAIRification of Scanning Tuneling Microscopy Images," *Data Intelligence* vol. 5, no. 1, pp. 27-42, 2023, doi: 10.1162/dint_a_00164.

[213] K. Ito, Y. Ogawa, K. Yokota, S. Matsumura, T. Minamisawa, K. Suga, K. Shiba, Y. Kimura, A. Hirano-Iwata, Y. Takamura, T. Ogino, "Host cell prediction of exosomes using morphological features on solid surfaces analyzed by machine learning," *The Journal of Physical Chemistry B* vol. 122, no. 23, pp. 6224-6235, 2018, doi: 10.1021/acs.jpcb.8b01646.

[214] G. Wang, W. Li, M. Aertsen, J. Deprest, S. Ourselin, T. Vercauteren, "Aleatoric uncertainty estimation with test-time augmentation for medical image segmentation with convolutional neural networks," *Neurocomputing* vol. 338, pp. 34-45, 2019, doi: 10.1016/j.neucom.2019.01.103.

[215] A. Ghosh, B. G. Sumpter, O. Dyck, S. V. Kalinin, M. Ziatdinov, "Ensemble learning-iterative training machine learning for uncertainty quantification and automated experiment in atom-resolved microscopy," *npj Computational Materials* vol. 7, no. 1, pp. 100, 2021, doi: 10.1038/s41524-021-00569-7.

[216] A. Pattison, C. Pedroso, B. E. Cohen, W. Theis, P. Ercius, "Advanced Techniques in Automated High Resolution Scanning Transmission Electron Microscopy," 2023 arXiv preprint arXiv:230305543.

[217] G. Wrzesiński, A. Markiewicz, "Prediction of Permeability Coefficient k in Sandy Soils Using ANN," *Sustainability* vol. 14, no. 11, pp. 6736, 2020, doi: 10.3390/su14116736.

[218] M. J. Miyama, K. Hukushima, "Real-space analysis of scanning tunneling microscopy topography datasets using sparse modeling approach," *Journal of the Physical Society of Japan* vol. 87, no. 4, pp. 044801, 2018, doi: 10.7566/JPSJ.87.044801.

[219] S. A. Meldgaard, H. L. Mortensen, M. S. Jørgensen, B. Hammer, "Structure prediction of surface reconstructions by deep reinforcement learning," *Journal of Physics: Condensed Matter* vol. 32, no, 40, pp. 404005, 2020, doi: 10.1088/1361-648X/ab94f2.

[220] Y. Liu, R. K. Vasudevan, K. P. Kelley, H. Funakubo, M. Ziatdinov, S. V. Kalinin, "Learning the right channel in multimodal imaging: automated experiment in piezoresponse force microscopy," *npj Computational Materials* vol. 9, no. 1, pp. 34, 2023, https://doi.org/10.48550/arXiv.2207.03039.

[221] D. M. Packwood, T. Hitosugi, "Rapid prediction of molecule arrangements on metal surfaces via Bayesian optimization," *Applied Physics Express* vol. 10, no. 6, pp. 065502, 2017, doi: 10.7567/APEX.10.065502.

[222] N. Käming, A. Dawid, K. Kottmann, M. Lewenstein, K. Sengstock, A. Dauphin, C. Weitenberg, "Unsupervised machine learning of topological phase transitions from experimental data," *Machine Learning: Science and Technology* vol. 2, no. 3, pp. 035037, 2021 doi: 10.1088/2632-2153/abffe7.

[223] D. V. Carvalho, E. M. Pereira, J. S. Cardoso, "Machine learning interpretability: A survey on methods and metrics," *Electronics* vol. 8, no. 8, pp. 832, 2019, doi: 10.3390/electronics8080832.

[224] M. Ziatdinov, U. Fuchs, J. H. Owen, J. N. Randall, S. V. Kalinin, "Robust multi-scale multi-feature deep learning for atomic and defect identification in Scanning Tunneling Microscopy on H-Si (100) 2x1 surface, 2020, arXiv preprint arXiv:200204716.

[225] M. Ragone, R. Shahabazian-Yassar, F. Mashayek, V. Yurkiv, "Deep learning modeling in microscopy imaging: A review of materials science applications," *Progress in Materials Science* vol. 138, pp. 101165, 2023, doi: 10.1016/j.pmatsci.2023.101165.

[226] F. Valdez, P. Melin, "A review on quantum computing and deep learning algorithms and their applications," *Soft Comput.* vol. 27, pp. 13217–13236, 2023, doi: 101007/s00500-022-07037-4.

[227] Y. Zhu, K. Yu, "Artificial intelligence (AI) for quantum and quantum for AI," *Opt. Quant. Electron.* vol. 55, pp. 697, 2023, doi: 101007/s11082-023-04914.

[228] R. Harishankar, J. Schaefer, M. Osborne, S. Muppidi, W. Rjaibi, "Security in the quantum computing era IBM Institute for Business Value," 2023, https://wwwibmcom/downloads/cas/EZEGKEB5.

[229] F. Hui, M. Lanza, "Scanning probe microscopy for advanced nanoelectronics," *Nat. Electron.* vol. 2, pp. 221–229, 2019, doi: 10.1038/s41928-019-0264-8.

[230] H. Wang, D. Lee, L. Wei, "Toward the Next Frontiers of Vibrational Bioimaging," *Chem. Biomed. Imaging* vol. 1, no. 1, pp. 3–17, 2023, doi: 10.1021/cbmi.3c00004.

[231] O. M. Gordon, P. J. Moriarty, "Machine learning at the (sub)atomic scale: next generation scanning probe microscopy," *Mach. Learn.: Sci. Technol.* vol. 1, no. 2, pp. 023001, 2020, doi: 10.1088/2632-2153/ab7d2f.

[232] Y. Zhang, A. Mesaros, K. Fujita, S. D. Edkins, M. H. Hamidian, K. Ch'ng, H. Eisaki, S. Uchida, J. C. S. Davis, E. Khatami, E. A. Kim, "Machine learning in electronic-quantum-matter imaging experiments," *Nature* vol. 570, pp. 484–490, 2019, doi: 10.1038/s41586-019-1319-8.

[233] R. M. Patton, J. T. Johnston, S. R. Young, C. D. Schuman, D. D. March, T. E. Potok, D. C. Rose, S. H. Lim, T. P. Karnowski, M. A. Ziatdinov, "167-PFlops deep learning for electron microscopy: from learning physics to atomic manipulation," in *Proc. Int. Conf. High Perform. Comput. Netw. Storage Anal.* IEEE Press 50, 2018, doi: 10.1109/SC.2018.00053.

[234] Y. Liu, R. K. Vasudevan, K. P. Kelley, H. Fudnakubo, M. Ziatdinov, S. V. Kalinin, "Learning the right channel in multimodal imaging: automated experiment in piezoresponse force microscopy," *npj Comput. Mater.* vol. 9, pp. 34, 2023, doi: 0.48550/arXiv.2207.03039.

[235] M. AlQuraishi, P. K. Sorger, "Differentiable biology: using deep learning for biophysics-based and data-driven modeling of molecular mechanisms," *Nat. Methods* vol. 18, no. 10, pp. 1169-1180, 2021, doi.org/10.1038/s41592-021-01283-4.

[236] A. Ghosh, B. Nachman, D. Whiteson, "Uncertainty-aware machine learning for high energy physics," *Phys. Rev. D* vol. 104, pp. 056026, 2021, doi: 10.1103/PhysRevD.104.056026.

[237] H. Bai, S. Wu, "Nanowire Detection in AFM Images Using Deep Learning," *Microscopy and Microanalysis* vol. 27, pp. 54–64, 2021, doi: 10.1017/S143192762002468X.

[238] S. Wang, J. Zhu, R. Blackwell, F. R. Fischer, "Automated Tip Conditioning for Scanning Tunneling Spectroscopy," *J. Phys. Chem. A* vol. 125, no. 6, pp. 1384–1390, 2021, doi: 10.1021/acs.jpca.0c10731.

[239] M. Ge, F. Su, Z. Zhao, D. Su, "Deep learning analysis on microscopic imaging in materials science," *Materials Today Nano* vol. 11, pp. 100087, 2020, doi: 10.1016/j.mtnano.2020.100087.

[240] J. C. Thomas, A. Rossi, D. Smalley, L. Francaviglia, Z. Yu, T. Zhang, S. Kumari, J. A. Robinson, M. Terrones, M. Ishigami, E. Rotenberg, E. S. Barnard, A. Raja, E. Wong, D. F. Ogletree, M. N. Noack, A. Weber-Bargioni, "Autonomous scanning probe microscopy investigations over WS2 and Au{111}," *npj Comput. Mater.* vol. 8, pp. 99, 2022, doi: 10.1038/s41524-022-00777-9.

[241] A. E. Gongora, V. Saygin, K. L. Snapp, K. A. Brown, "Autonomous experimentation in nanotechnology," in *Materials Today* pp. 331-360, Elsevier 2023.

[242] H. Li, J. Jiao, K. Davey, S. Z. Qiao, "Data-Driven Machine Learning for Understanding Surface Structures of Heterogeneous Catalysts," *Angewandte Chemie* vol. 135, no. 9, pp. e(2022)16383, 2022, doi: 10.1002/anie.202216383.

[243] L. Yao, Q. Chen, "Machine learning in nanomaterial electron microscopy data analysis," in Materials Today Intelligent Nanotechnology, Chapter 10, pp. 279-305.

[244] K. Choudhary, R. Gurunathan, B. DeCost, A. Biacchi, "AtomVision: A Machine Vision Library for Atomistic Images," *J. Chem. Inf. Model.* vol. 63, no. 6, pp. 1708–1722, 2023 doi: 10.1021/acs.jcim.2c01533.

[245] S. V. Kalinin, R. K. Vasudevan, Y. Liu, A. Ghosh, K. Raccapriore, M. Ziatdinov, "Probe microscopy is all you need," *Mach. Learn.: Sci. Technol.* vol. 4, no. 2, pp. 023001, 2023, doi.org/101088/2632-2153/acccd5.

[246] X. Peng, X. Wang, "Next-generation intelligent laboratories for materials design and manufacturing," *MRS Bulletin* vol. 48, pp. 179–185, 2023, doi: 10.1557/s43577-023-00481-z.

[247] A. Dujardin, P. De Wolf, F. Lafont, V. Dupres, "Multi-Sample Acquisition and Analysis Using Atomic Force Microscopy for Biomedical Applications," *PLoS ONE* vol. 14, pp. e0213853, 2019, doi: 10.1371/journal.pone.0213853.

[248] W. K. Szeremeta, R. L. Harniman, C. R. Bermingham, M. Antognozzi, "Towards a Fully Automated Scanning Probe Microscope for Biomedical Applications," *Sensors* vol. 2621, no. 9, pp. 3027, 2021, doi: 10.3390/s21093027.

[249] V. Tshitoyan, J. Dagdelen, L. Weston, A. Dunn, Z. Rong, O. Kononova, K. A. Persson, G. Ceder, A. Jain, "Unsupervised word embeddings capture latent knowledge from materials science literature," *Nature* vol. 571, pp. 95–98, 2019, doi: 10.1038/s41586-019-1335-8.

[250] J. M. Gregoire, L. Zhou, J. A. Haber, "Combinatorial synthesis for AI-driven materials discovery," *Nat. Synth.* pp. 1-12, 2023, doi: 10.1038/s44160-023-00251-4.

[251] J. Qin, B. Sun, G. Zhou, T. Guo, Y. Chen, C. Ke, S. Mao, X. Chen, J. Shao, Y. Zhao," *ACS Materials Letters* vol. 5, no. 8, pp. 2197-2215, 2023, doi: 101021/acsmaterialslett3c00088.

[252] A. Melnikov, M. Kordzanganeh, A. Alodjants, R. K. Lee, "Quantum machine learning: from physics to software Engineering," *Advances in Physics: X* vol. 8, pp. 1, 2023, doi: 101080/2374614920232165452.